\documentclass[%
 reprint,
nofootinbib,
 amsmath,amssymb,
 aps,
]{revtex4-1}

\usepackage{graphicx}
\usepackage[dvipsnames]{xcolor}
\usepackage{dcolumn}
\usepackage{bm}
\usepackage[colorlinks]{hyperref}

\usepackage{mathrsfs}

\newcommand\myshade{80}
\colorlet{mylinkcolor}{ForestGreen}
\colorlet{mycitecolor}{Red}
\colorlet{myurlcolor}{violet}

\hypersetup{
  linkcolor  = mylinkcolor!\myshade!black,
  citecolor  = mycitecolor!\myshade!black,
  urlcolor   = myurlcolor!\myshade!black,
  colorlinks = true
}

\begin{document}

\preprint{APS/123-QED}

\newcommand{\GRAPPA}{%
Gravitation Astroparticle Physics Amsterdam (GRAPPA),\\
Institute for Theoretical Physics Amsterdam
and Delta Institute for Theoretical Physics,\\
University of Amsterdam, Science Park 904, 1098 XH Amsterdam, The Netherlands\\}

\newcommand{\CHALMERS}{
Chalmers University of Technology,\\
Department of Physics,\\
SE-412 96 G\"oteborg, Sweden
}

\newcommand{\IFCA}{Instituto de F\'isica de Cantabria (IFCA, UC-CSIC), Av.~de
Los Castros s/n, 39005 Santander, Spain}

\title{Measuring the local Dark Matter density in the laboratory}

\author{Bradley J. Kavanagh}
\email{kavanagh@ifca.unican.es}
\affiliation{\IFCA}
\affiliation{\GRAPPA}

\author{Timon Emken}%
 \email{emken@chalmers.se}
\affiliation{\CHALMERS}

\author{Riccardo Catena}%
 \email{catena@chalmers.se}
\affiliation{\CHALMERS}

\date{\today}

\begin{abstract}
Despite strong evidence for the existence of large amounts of dark matter~(DM) in our Universe, there is no direct indication of its presence in our own solar system.
All estimates of the local DM density rely on extrapolating results on much larger scales.
We demonstrate for the first time the possibility of simultaneously measuring the local DM density and interaction cross-section with a direct detection experiment.
It relies on the assumption that incoming DM~particles frequently scatter on terrestrial nuclei prior to detection, inducing an additional time-dependence of the signal.
We show that for sub-GeV DM, with a large spin-independent DM-proton cross section, future direct detection experiments should be able to reconstruct the local DM~density with smaller than 50\% uncertainty.
\end{abstract}

\maketitle


\section{Introduction}
A self-gravitating fluid that does not emit or absorb radiation at any observable wavelength, Dark Matter (DM) is the only coherent explanation for a number of otherwise anomalous phenomena~\cite{Bertone:2004pz,Bertone:2016nfn}.~These range from stellar motions in nearby dwarf spheroidal galaxies~\cite{Lokas2002Jul} to anisotropies in the cosmic microwave background radiation~\cite{Ade:2015xua}.~There is also strong evidence for the presence of DM in the Milky Way (MW), as inferred from kinematic measurements of stellar populations~\cite{Brown:2018dum}, microlensing events~\cite{Moniez:2010zt} and the dynamics of satellite galaxies~\cite{Callingham:2018vcf}.

While the evidence for DM in the Universe and in our own Galaxy is compelling, there is no {\it direct} indication of the presence of DM within about one parsec of the Sun~\cite{Tremaine1990}.~The only available estimates divide into two classes:~1) local methods based on the vertical motion of stellar populations~\cite{Kuijken:1989hu,MoniBidin2012,Bovy:2012tw,Garbari2012,Smith2012,Zhang2013,Bovy:2013raa,Bienayme:2014kva,McKee:2015hwa,Xia:2015agz,Sivertsson:2017rkp,Buch:2018qdr}; 2) global methods relying on mass models for the MW~\cite{Salucci:2010qr,Catena:2009mf,Weber2010,Iocco:2011jz,McMillan2010,Pato:2015dua,Huang2016,McMillan2017,2019JCAP...10..037D}.~Each method comes with its own limitations as well as systematic and statistical errors~\cite{Peter:2011eu,Fairbairn:2012zs,Kavanagh:2012nr,Kavanagh:2013eya,Kavanagh:2013wba,Kavanagh:2014rya,Benito:2019ngh,2019JCAP...09..046K}.~However, no currently known astronomical tracer can directly probe the DM contribution to the MW gravitational potential with sub-parsec resolution~\cite{2013AstL...39..141P,Read:2014qva}.

Progress in understanding the particle properties of DM~\cite{Peter:2012jh,Fan:2013tia,Fan:2013yva}, the shape, composition and merger history  of  the  MW~\cite{Read:2008fh,10.1111/j.1365-2966.2009.14757.x} and, more broadly, the formation of galaxies is hampered by the lack of such ultra-local sub-parsec information about the DM density. In particular, by combining global and ultra-local measurements of the DM density, we can constrain the shape of the MW halo~\cite{Read:2014qva}. This in turn may resolve a long-standing tension in standard $\Lambda$CDM cosmology: theory and simulations predict triaxial DM halos~\cite{Dubinski:1993df,Kazantzidis:2004vu,Maccio:2006wpz,Debattista_2008,Hellwing:2011ne,2019MNRAS.484..476C}, while observations in the MW~\cite{Ibata:2000pu,2010ApJ...714..229L,Bovy:2016chl,Wegg2019May} point towards a roughly spherical halo. Although significant effort has recently been made to explain the observed halo shapes with hydrodynamic simulations~\cite{Dai:2018ypv,Prada2019Dec,Poole-McKenzie:2020dbo,Cataldi:2020kvs,Emami:2020cwt}, a direct measurement of the local DM density would provide a crucial, independent test of our understanding of Galaxy formation, with important implications for astronomy, astrophysics, cosmology and particle physics.

The lack of direct astronomical measurements of the DM density at the Earth's location also hinders the success of terrestrial `direct detection' experiments~\cite{Drukier:1983gj,Goodman:1984dc,Drukier:1986tm}.~These detectors search for DM-nucleus scattering events in underground laboratories, with an expected event rate depending on both the local DM density and the DM-nucleus scattering cross section.

Here, we explore a radically new approach to the problem of finding the local DM density at the Earth's location.~We propose to exploit the diurnal variation of the DM flux
{\it after} Earth-crossing to {\it simultaneously} measure the local DM density $\rho_\chi$ and DM-nucleus scattering cross section $\sigma$ with future direct detection experiments.~This diurnal variation arises from distortions in the DM distribution, due to interactions of DM particles in the Earth before they reach the detector~\cite{Gould:1988eq,Collar:1993ss,Hasenbalg:1997hs}. The amplitude of this modulation depends on the cross section~\cite{Kavanagh:2016pyr,Emken:2017qmp}, as we will demonstrate via Monte Carlo (MC) simulations, allowing us to break the degeneracy between $\rho_\chi$ and $\sigma$. A similar method has recently been used to measure the high-energy neutrino-nucleon cross section with IceCube~\cite{Aartsen:2017kpd}. We show that using event timing information, combined with the energy spectrum of a hypothetical DM signal, can enable a measurement of the local DM density and cross-section with low-threshold experiments.  

Throughout this paper, we emphasise the DM density measurement, given that our proposal is potentially the only method of directly pinning down $\rho_\chi$.~\footnote{For a related proposal, with DM leaving multi-scatter tracks in the detector, see Ref.~\cite{Bramante:2018qbc}.}
We find that the precision of this measurement depends on  the detector's location
and can be smaller than about 50\% for DM-proton scattering cross sections larger than 10$^{-32}$~cm$^2$ and a DM mass around 100 MeV, becoming much more precise for larger cross sections.~Here, we focus on DM-nucleus scattering, but, if extended to DM-electron interactions~\cite{Essig:2011nj} or more exotic detection strategies (e.g.~\cite{Knapen:2016cue,Hochberg:2017wce,Trickle:2019ovy,Hochberg:2019cyy}), our method can be applied to DM candidates in the keV to sub-GeV range, covering a significant fraction of the parameter space of detectable DM candidates~\cite{Bertone:2018xtm}. 

The associated simulation and statistics codes are publicly available at~\cite{Emken2017} and~\cite{LikelihoodCode} respectively.

\section{Direct detection formalism} 

The differential recoil rate for a DM particle of mass~$m_\chi$ with a nucleus~$A$ of mass~$m_A$ can be written~\cite{Lewin:1995rx,Cerdeno:2010jj}
\begin{align}
\frac{\mathrm{d}R}{\mathrm{d}E_R} = \frac{\rho_\chi}{ m_\chi m_A} \int_{v > v_\mathrm{min}}\mathrm{d}^3\mathbf{v}\; v\, f(\mathbf{v}) \frac{\mathrm{d}\sigma^\mathrm{SI}}{\mathrm{d}E_R}\,, \label{eq: nuclear recoil spectrum}
\end{align}
with local DM density $\rho_\chi$ and local DM~velocity distribution in the laboratory~$f(\mathbf{v})$. Neglecting the effect of Earth scatterings, the usual choice for $f(\mathbf{v})$ in the context of direct detection is the \textit{standard halo model}~(SHM)~\cite{Lewin:1995rx,Green:2011bv,Green:2017odb}, a Maxwell-Boltzmann distribution in the galactic frame, truncated at the local galactic escape speed~$v_{\rm esc}\approx 544\,\text{km s}^{-1}$~\cite{Smith:2006ym,Piffl:2013mla}.
We integrate over $v > v_\mathrm{min}$, the minimum speed kinematically required to produce a nuclear recoil of energy $E_R$. It is a crucial feature of this work that for large enough cross sections both $\rho_\chi$ and  $f(\mathbf{v})$ are modified by underground scatterings, thereby modifying the rate in Eq.~\eqref{eq: nuclear recoil spectrum}. 

The true recoil energy~$E_R$ does not directly correspond to the detected energy deposit~$E_D$.
We account for a finite energy resolution by transforming the theoretical recoil spectrum of Eq.~\eqref{eq: nuclear recoil spectrum} into the observed spectrum
\begin{align}
    \frac{\mathrm{d}R}{\mathrm{d}E_D} = \int_{E_R^\mathrm{min}}^\infty \mathrm{d} E_R\, \text{Gauss}(E_D|\mu=E_R,\sigma_E) \frac{\mathrm{d}R}{\mathrm{d}E_R} \, . \label{eq: nuclear recoil spectrum detected}
\end{align}
Here, we model the detector response function as a Gaussian with mean~$E_R$ and standard deviation or energy resolution~$\sigma_E$.
For a given energy threshold~$E_\mathrm{th}$, a finite energy resolution means that a nuclear recoil below the threshold, $E_R<E_\mathrm{th}$ might fluctuate above the threshold and be detectable.
However, the approximation of a Gaussian breaks down for energies far below the threshold~\cite{Angloher:2017zkf}, which is why we set~$E_R^\mathrm{min} = E_\mathrm{th}-2\sigma_E$ and thereby only include up-fluctuations of $2\sigma$ to avoid unphysical signal rates. 

We also assume standard spin-independent (SI) interactions for the differential scattering cross section,
\begin{align}
\frac{\mathrm{d}\sigma^\mathrm{SI}}{\mathrm{d}E_R} =  \frac { m_{A} \sigma^{\mathrm{SI}}_{p}} { 2 \mu _ { \chi p } ^ { 2 } v ^ { 2 } } A ^ { 2 } F ^ { 2 } \left( E _ { R } \right)\,. \label{eq: differential SI cross section}
\end{align}
Here, $\sigma^\mathrm{SI}_p$ is the DM-proton cross section at zero momentum transfer and $A$ the nucleus' mass number. We consider light DM, $m_\chi \ll m_A$, so we set the nuclear form factor $F^2(E_R) = 1$. 

\begin{figure*}
    \centering
    \includegraphics[width=0.3\textwidth]{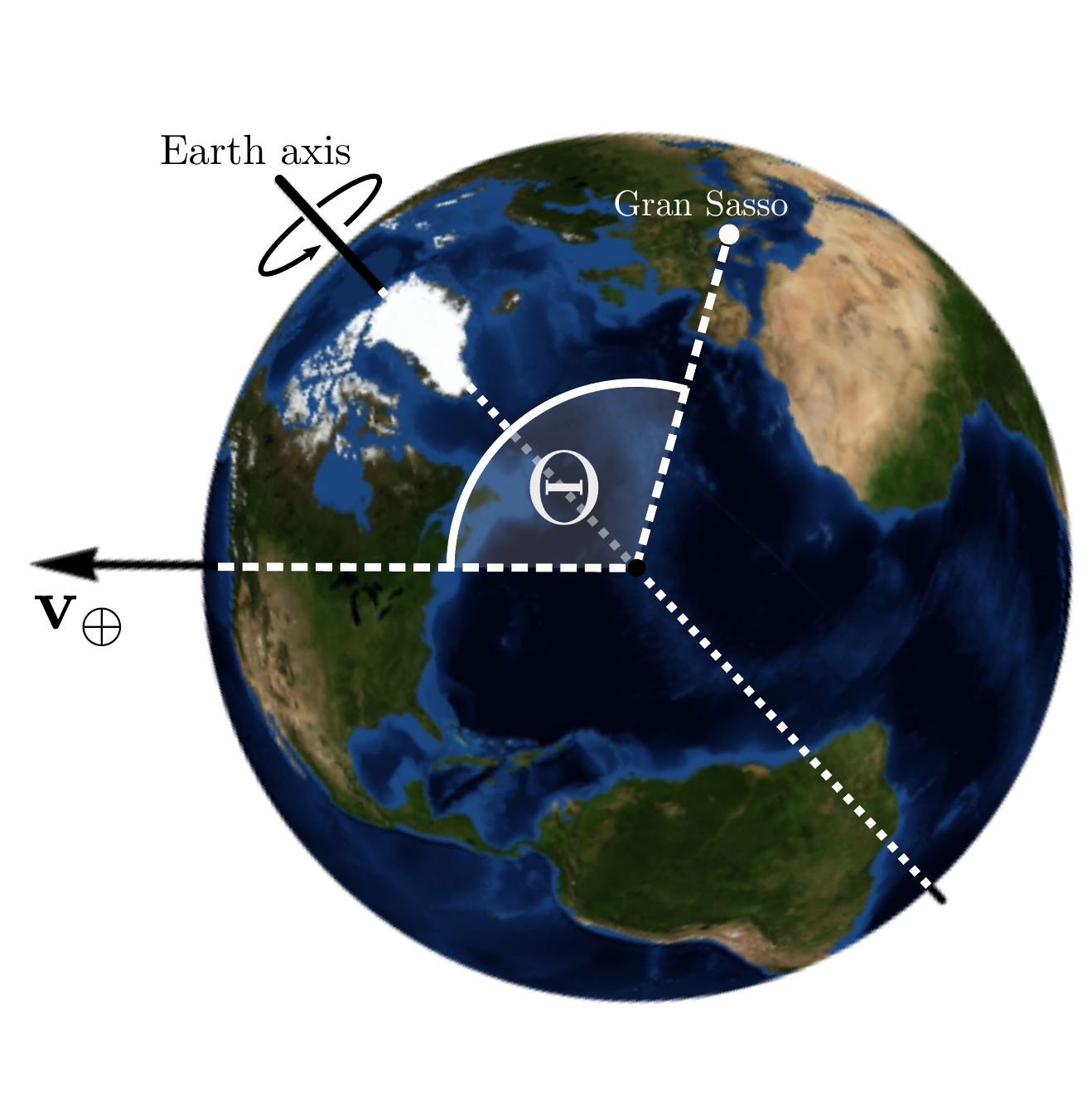}
    \includegraphics[width=0.65\textwidth]{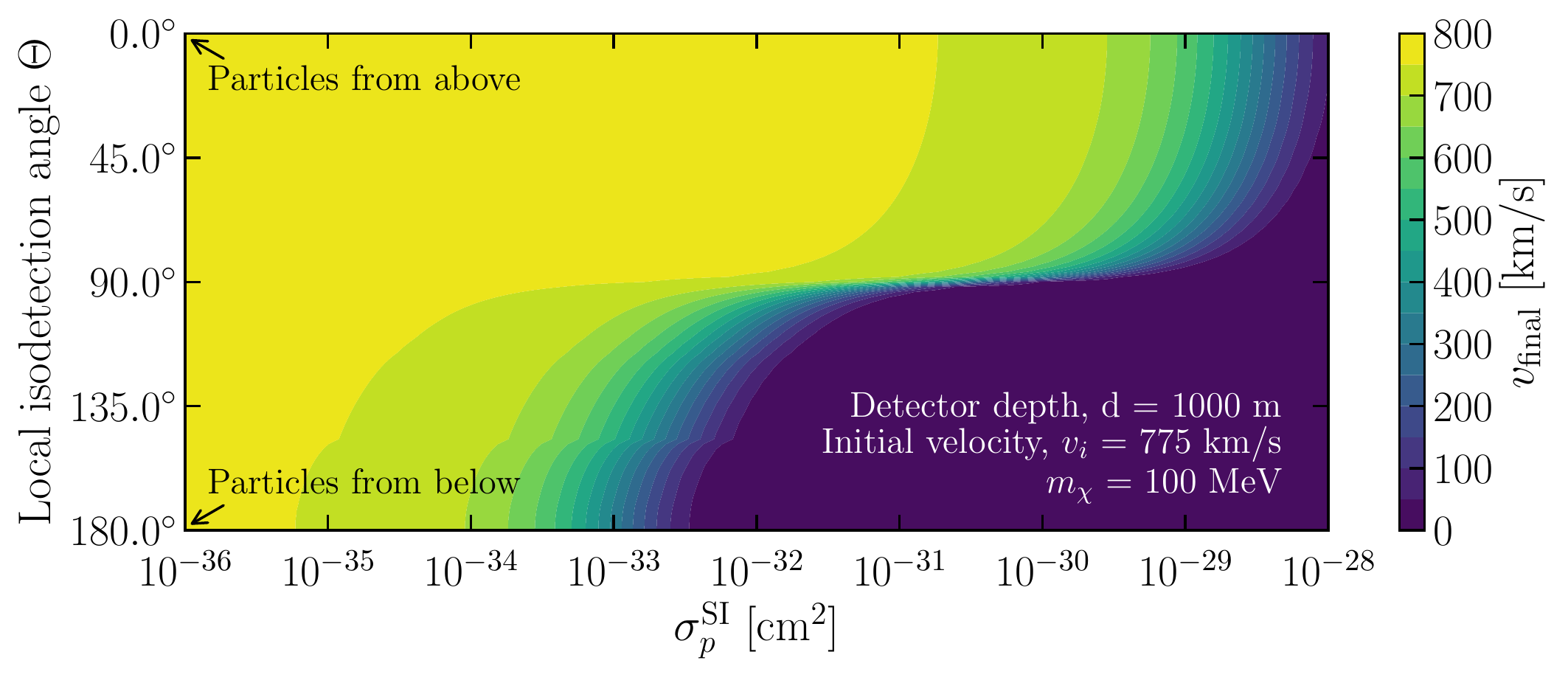}
    \caption{\textbf{Left:} Visualization of the local isodetection angle defined in Eq.~\eqref{eq: isodetection angle} at LNGS and a specific moment in time. \textbf{Right:} Final velocity of DM particles as a function of isodetection angle and DM-proton cross section $\sigma_p^\mathrm{SI}$. For illustration only, we assume straight-line trajectories of DM particles~\cite{Kavanagh:2017cru} with initial speed $v_\oplus + v_\mathrm{esc} \approx 775 \,\mathrm{km/s}$, travelling in the mean direction of the DM flux $-\mathbf{v}_\oplus$ (left-to-right in the left panel).}
    \label{fig: isodetection angle}
\end{figure*}

While we focus on spin-independent interactions as a proof of concept, similar analyses could just as well be performed for spin-dependent scattering \cite{Cerdeno:2010jj}, long-range interactions \cite{Fornengo:2011sz,DelNobile:2012tx,Kahlhoefer:2017ddj} or the broader class of effective field theory interactions \cite{Fan:2010gt,Fitzpatrick:2012ix,DelNobile:2013sia,DelNobile:2018dfg,Catena:2019hzw}. Indeed, similar results should also apply for DM-electron scattering \cite{Essig:2011nj,Essig:2015cda,Derenzo:2016fse}.

\section{Earth scattering}

Above a certain DM-proton cross section~$\sigma^{\rm SI}_p \gtrsim 10^{-37}\,\text{cm}^2$, the probability for a DM~particle to scatter on a terrestrial target
becomes non-negligible. In this regime, underground scatterings prior to passing through the detector decelerate and deflect the incoming DM~particles and thereby change the local DM~density and distribution. These distortions grow with the cross section, and the signal thus depends non-linearly on $\sigma_p^{\rm SI}$.

In the single-scattering regime of moderate cross sections, the impact of Earth scatterings on the local DM~properties can be quantified analytically~\cite{Kavanagh:2016pyr}.
However, the precise contributions of multiple scatterings require the use of MC simulations of underground DM~particle trajectories~\cite{Collar:1993ss,Hasenbalg:1997hs,Emken:2017qmp}, where we use the numerical tool \textsc{DaMaSCUS}~\cite{Emken2017}\footnote{Similar MC~simulations have been used to study the sensitivity of terrestrial experiments to strongly interacting~DM~\cite{Zaharijas:2004jv,Emken:2017erx,Mahdawi:2017cxz,Emken:2018run,Mahdawi:2018euy,Emken:2019tni}. However, a number of analytic approximations have also been applied in this context~\cite{Albuquerque:2003ei,Kouvaris:2014lpa,Kouvaris:2015laa,Davis:2017noy,Kavanagh:2017cru,Hooper:2018bfw,Bramante:2018qbc,Bramante:2018tos,Bramante:2019yss}.}.
The simulation details are described extensively in~\cite{Emken:2017qmp,Emken:2019hgy}, and we briefly review the essentials here.

The shape of a DM~particle's trajectory is primarily determined by the local mean free path,
\begin{align}
    \lambda^{-1}(\mathbf{x}) &= \sum_i \lambda_i^{-1}(\mathbf{x})\textbf{}\equiv\sum_i n_i(\mathbf{x})\sigma^{\rm SI}_i\, ,
\end{align}
where $n_i(\mathbf{x})$ is the local number density of isotope~$i$, and~$\sigma^{\rm SI}_i$ is the total DM-nucleus scattering cross section for that nucleus.
The number densities depend on the Earth's mass density profile~$\rho_\oplus(r)$, taken from the Preliminary Reference Earth Model~(PREM)~\cite{Dziewonski:1981xy}, and the relative nuclear abundances~\cite{McDonough:2003}.
Furthermore, the distribution of the DM-nucleus scattering angle~$\theta$ arises from the differential cross section in Eq.~\eqref{eq: differential SI cross section} and the relation between~$\theta$ and the recoil energy,~$E_R \propto (1-\cos\theta)$.

The simulated system features an axial symmetry around the direction of the Earth's velocity~$\mathbf{v}_\oplus$. 
This symmetry allows us to define the \textit{isodetection angle}~$\Theta$~\cite{Collar:1993ss,Hasenbalg:1997hs}, the polar angle from the symmetry axis as illustrated in Fig.~\ref{fig: isodetection angle}.
The time-dependent local isodetection angle of a terrestrial observer at~$\mathbf{x}_\mathrm{obs}$ reads
\begin{align}
    \Theta \equiv \angle(\mathbf{v}_\oplus,\mathbf{x}_\mathrm{obs}) =  \arccos \left[ \frac{\mathbf{v}_\oplus\cdot \mathbf{x}_{\rm obs}}{v_\oplus (r_\oplus-d)}\right]\, , \label{eq: isodetection angle}
\end{align}
where~$r_\oplus \approx  6370\,\mathrm{km}$ is the Earth's radius, and $d\sim 1\text{ km}$ is the underground depth of the observer.
It varies over a sidereal day, as described e.g. in App.~A of~\cite{Emken:2017qmp}.

To extract local estimates based on the MC~simulations, we define isodetection rings of finite size~$\Delta\Theta=5^\circ$.
By counting the particles passing through each isodetection ring, we obtain an MC estimation of the local DM~density~$\hat{\rho}_\chi$.
By recording their speeds, we obtain a (weighted) histogram estimate of the local speed distribution~$\hat{f}(v,\Theta)$~\cite{Emken:2017qmp}.
Finally, these estimates are used to determine the local nuclear recoil spectrum expected for a given value of~$\Theta$ via Eq.~\eqref{eq: nuclear recoil spectrum}. We performed a grid of 45 MC~simulations and evaluated the recoil spectra for DM~parameters in the ranges $m_\chi \in [0.058, 0.5]\,\mathrm{GeV}$ and $\sigma_p^\mathrm{SI} = [10^{-38}, 10^{-30}]\,\mathrm{cm}^2$, accounting for the crucial impact of Earth scatterings. Below $m_\chi \approx 0.058 \,\mathrm{GeV}$, the experimental setups we consider begin to rapidly lose sensitivity, due to the exponential suppression of events above the energy threshold. 

\section{Extracting the local DM density from data}
We express the sensitivity of direct detection experiments to the local DM density in terms of contours of constant $p$-value.~We can then reject a point $\boldsymbol{\theta}=(\sigma_p^{\rm SI}, \rho_\chi)$ on these contours in favour of the alternative, benchmark point $\boldsymbol{\theta}'=(\sigma_p^{\rm SI \prime}, \rho_\chi')$ with a statistical significance of $\Phi^{-1}(1-p)$, where $\Phi$ is the standard normal distribution.~For the local DM density, we assume $\rho_\chi'=0.4$~GeV~cm$^{-3}$.  

We calculate such $p$-value contours by using $t_{\boldsymbol{\theta}} = -2 \ln \lambda(\boldsymbol{\theta})$ as a test statistic, where $\lambda(\boldsymbol{\theta})$ is the profile likelihood ratio, defined in Eq.~(7) of~\cite{Cowan:2010js}.~We account for the unknown DM mass by maximising the likelihood (at fixed $\boldsymbol{\theta}$) with respect to $m_\chi \in [0.058, 0.5]\,\mathrm{GeV}$.~The $p$-value calculation requires the probability density function (pdf) of $t_{\boldsymbol{\theta}}$ under the assumption that the true model parameters are $\boldsymbol{\theta}$ or $\boldsymbol{\theta}'$.~We denote these pdfs by $f(t_{\boldsymbol{\theta}}|\boldsymbol{\theta})$ and $f(t_{\boldsymbol{\theta}}|\boldsymbol{\theta}')$, respectively.~Following~\cite{Cowan:2010js}, we approximate $f(t_{\boldsymbol{\theta}}|\boldsymbol{\theta})$ as a chi-square distribution with $k=2$ degrees of freedom and $f(t_{\boldsymbol{\theta}}|\boldsymbol{\theta}')$ as a non-central chi-square distribution with the same number of degrees of freedom~\cite{Cowan:2010js} and non-centrality parameter $\Lambda= - 2\ln \lambda (\boldsymbol{\theta})$.~Here, we restrict ourselves to ``Asimov data'', defined as the hypothetical dataset such that the maximum likelihood estimator, $\hat{\boldsymbol{\theta}}$, and benchmark point, $\boldsymbol{\theta}'$, coincide.~The $p$-value for rejecting the hypothesised point 
$\boldsymbol{\theta}$ in favour of $\boldsymbol{\theta}'$ is then given by
\begin{align}
p= \int_{t_{\boldsymbol{\theta}}>t_{\rm med}} {\rm d}t_{\rm \boldsymbol{\theta}} f(t_{\boldsymbol{\theta}}|\boldsymbol{\theta}) \,,
\label{eq:p}
\end{align}
where $t_{\rm med}$ is the median of $f(t_{\boldsymbol{\theta}}|\boldsymbol{\theta}')$.

The profile likelihood ratio, $\lambda(\boldsymbol{\theta})$, depends on the expected number of nuclear recoils from DM signal and background events in the $i$-th energy bin and in the $j$-th time bin, $s_{ij}$ and $b_{ij}$, respectively (see Eq.~(7) in~\cite{Cowan:2010js}).~We calculate $s_{ij}$, $i=1,\dots,N$, $j=1,\dots,M$, by integrating Eq.~(\ref{eq: nuclear recoil spectrum}) over $N=12$ ($M=12$) energy (time) bins of equal size.
We consider two experimental setups.
Motivated by existing experiments~\cite{Armengaud:2019kfj,Arnaud:2020svb}, the first detector we consider is a germanium detector, with a Gaussian energy resolution of 18~eV. The energy bins in the analysis cover the energy interval from the assumed threshold, 60 eV, to a maximum energy of 500 eV. 
The second detector we consider is a cryogenic calorimeter with a sapphire target ($\text{Al}_2\text{O}_3$), inspired by the~$\nu$-cleus experiment~\cite{Strauss:2017cam,Angloher:2017sxg}. 
For the energy threshold and resolution, we assume~$E_\mathrm{th}=10\text{ eV}$ and~$\sigma_E=3\text{ eV}$ respectively, which should be achievable for sapphire targets with some improvements in detector performance~\cite{Strauss:2017cam}. We assume perfect detection efficiency for both detectors.

The exposure spans a total of 30 days, starting from January 1st 2020, folded onto a single sidereal day, which is then divided into $M=12$ time bins. For both detectors, we assume a target mass of 35~g, leading to a total exposure of 1~kg~day.\footnote{Such a large target mass is unlikely to be possible with a single sapphire target, while preserving the low threshold of 10~eV~\cite{Strauss:2017cam}. However, it is conceivable that an array of gram-scale targets could be operated. In any case, we find that our results are background-limited rather than exposure-limited.}~We calculate $b_{ij}$ assuming a time-independent background consisting of a flat component and an exponentially falling component, as observed by EDELWEISS-Surf~\cite{Armengaud:2019kfj}. We assume that both detectors are operated at a depth $d = 1000\,\mathrm{m}$ underground.

We consider two benchmark masses for the DM particle. The first benchmark is $m_\chi' = 400 \,\mathrm{MeV}$, for which the DM-proton scattering cross section is constrained to be $\sigma_p^\mathrm{SI} \lesssim 10^{-37}\,\mathrm{cm}^2$ by current direct searches~\cite{Abdelhameed:2019hmk}. We consider searches for this particle with the germanium detector ($E_\mathrm{th} = 60\,\mathrm{eV}$). The second benchmark is $m_\chi' = 100 \,\mathrm{MeV}$, which is significantly less constrained: cross sections of $\sigma_p^\mathrm{SI} \lesssim 5 \times 10^{-31}\,\mathrm{cm}^2$ are still allowed by current constraints~\cite{Cappiello:2019qsw}. A very low threshold is required for sensitivity to such light DM and we therefore consider searches for this particle with the sapphire detector ($E_\mathrm{th} = 10\,\mathrm{eV}$).

\begin{figure*}[tbh]
    \centering
    \includegraphics[width=0.49\textwidth]{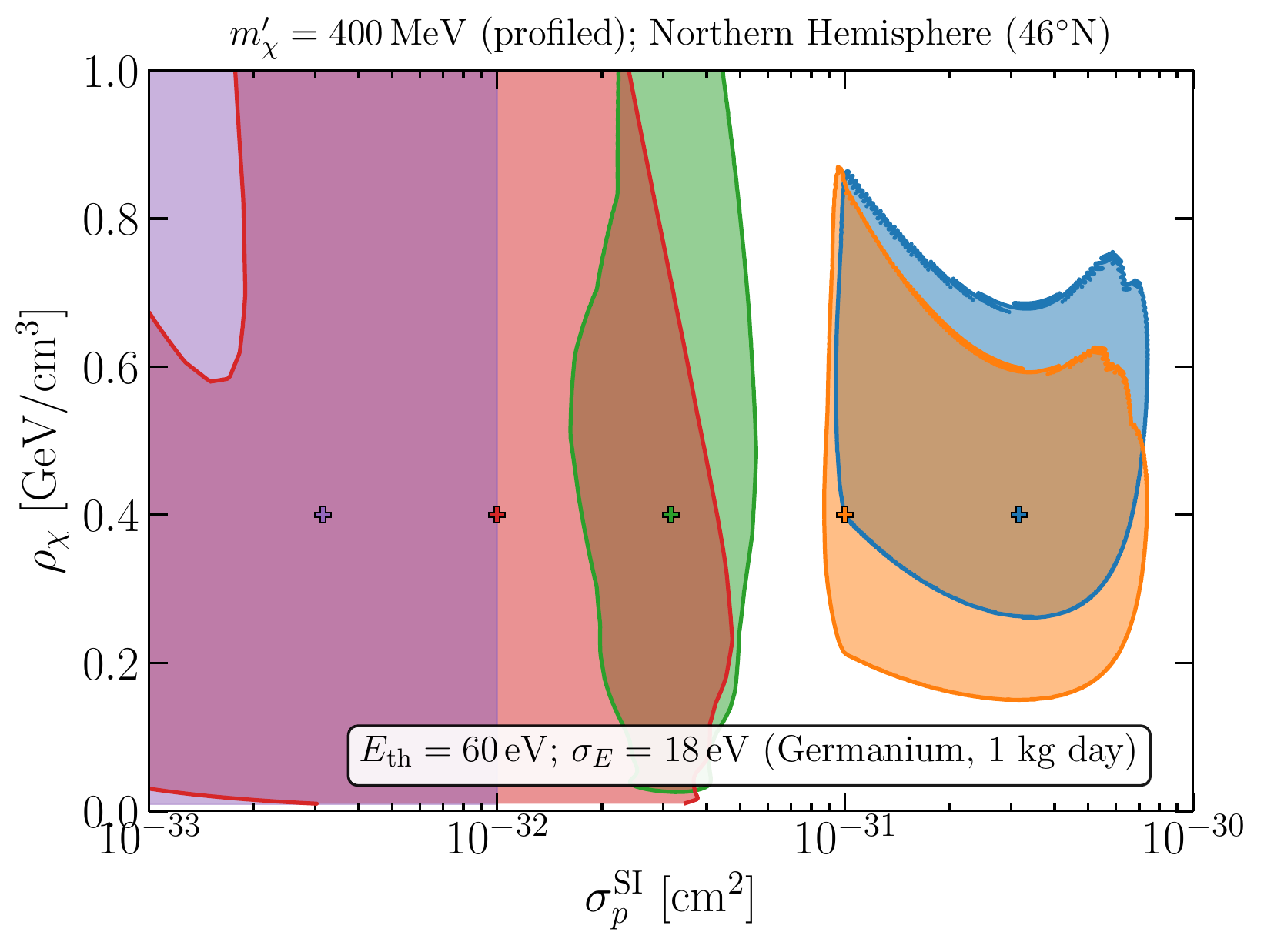}
    \includegraphics[width=0.49\textwidth]{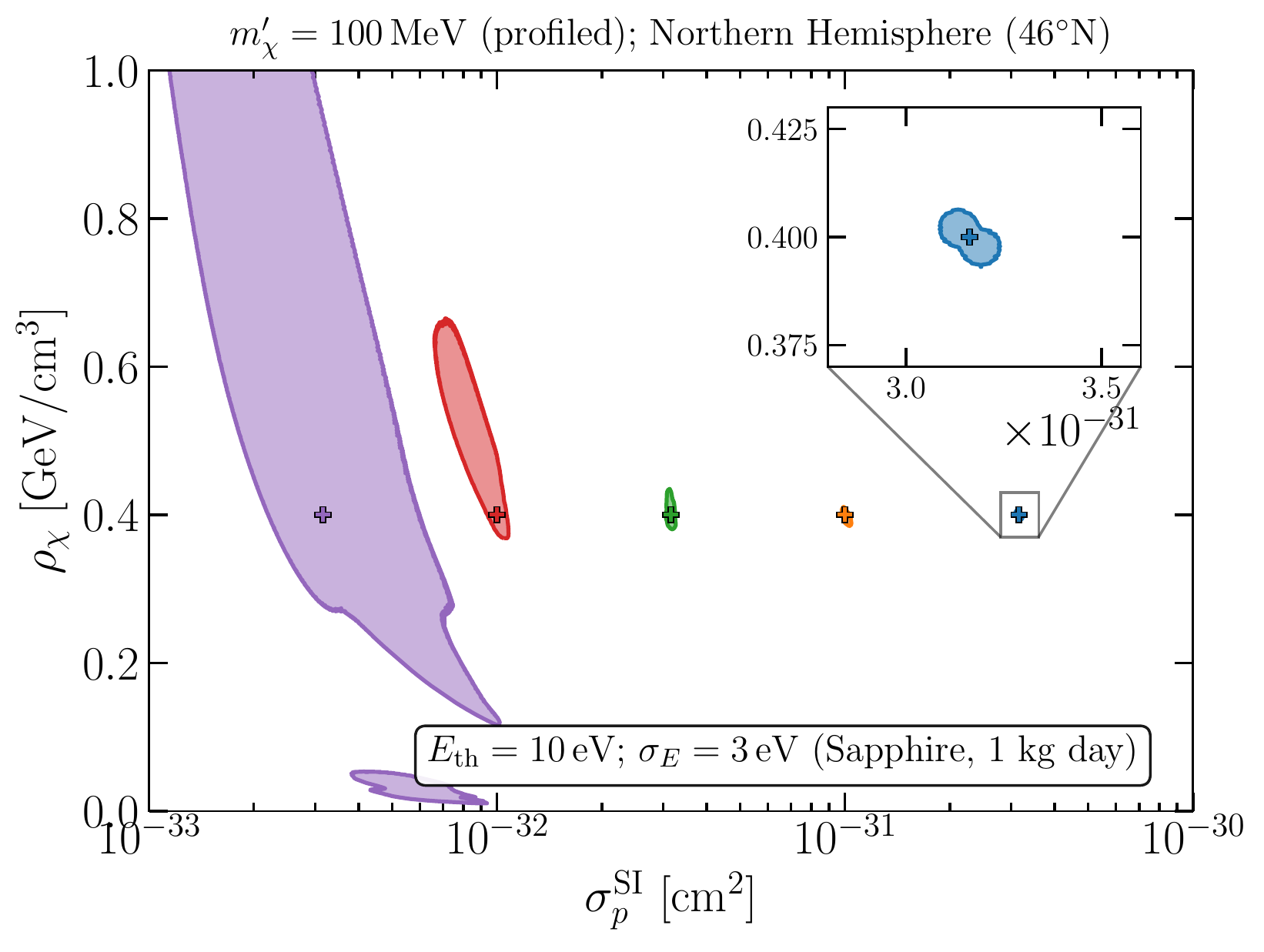}
    \includegraphics[width=0.49\textwidth]{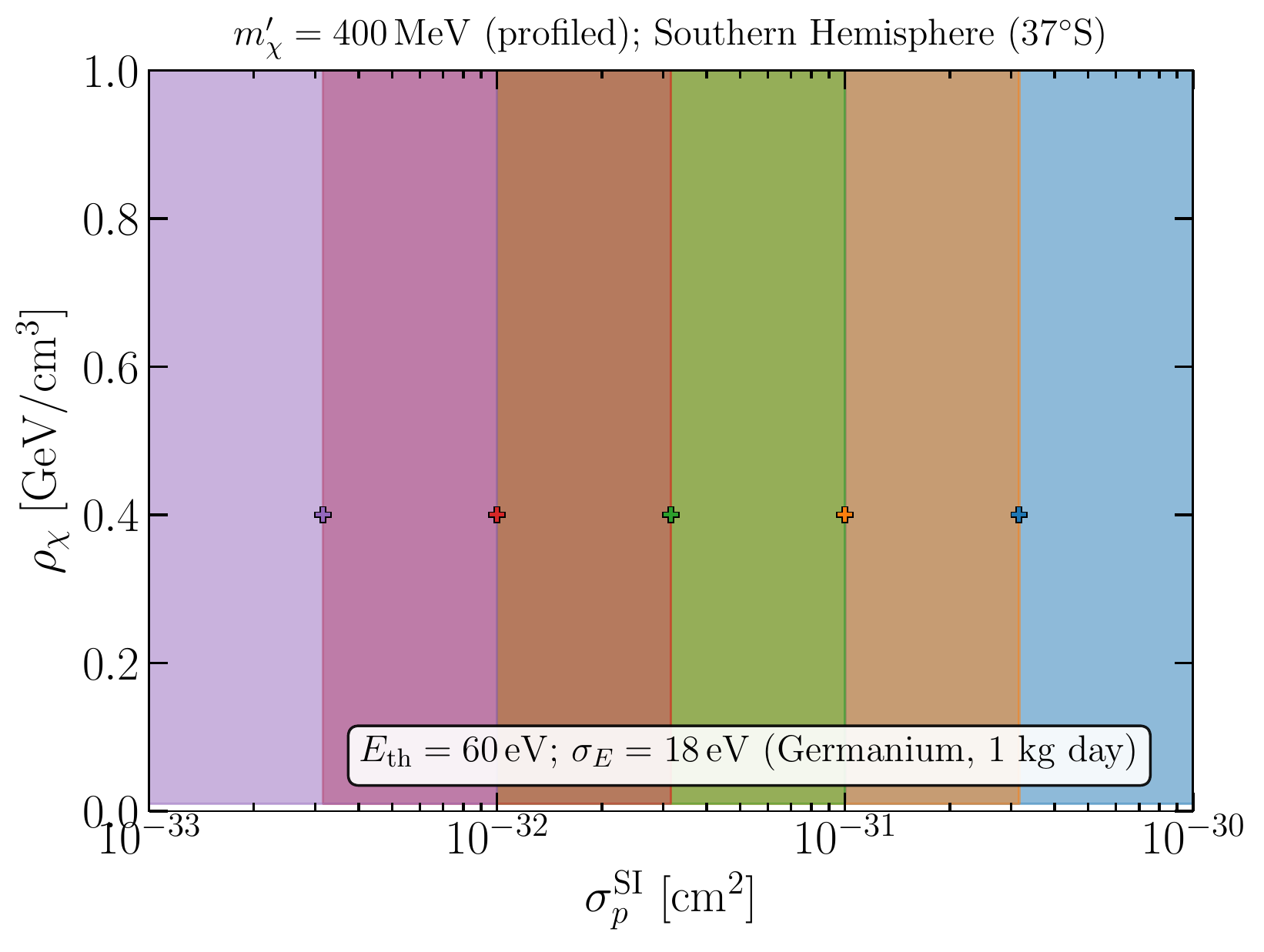}
    \includegraphics[width=0.49\textwidth]{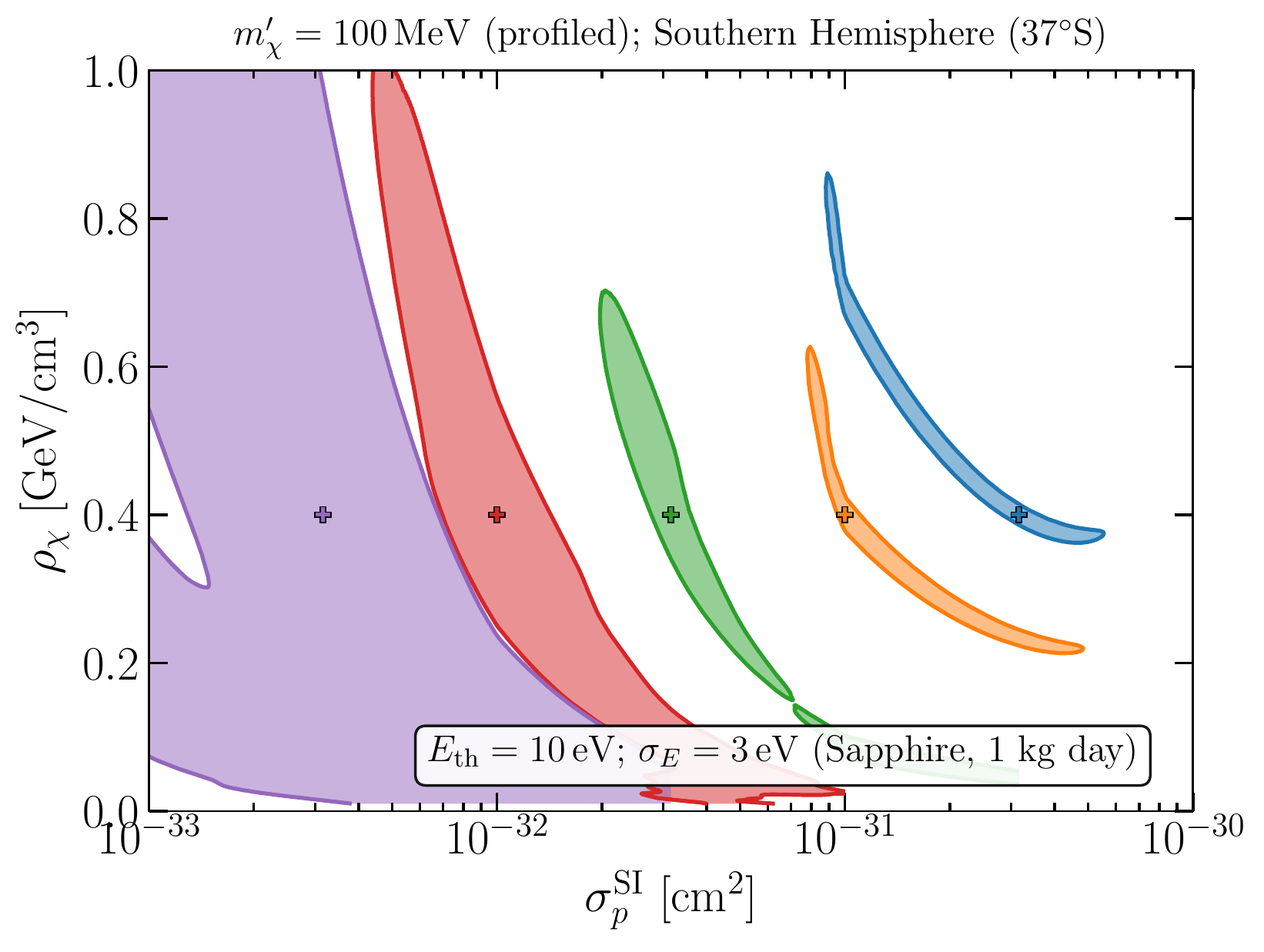}
    \caption{Projected 95\% CL contours using both energy and timing of events. In all cases, we assume a benchmark DM density of $\rho_\chi^\prime = 0.4 \,\mathrm{GeV}/\mathrm{cm}^3$. We consider 5 different benchmark cross sections, denoted by different colors. \textbf{Left:} Germanium detector and benchmark DM mass $m_\chi^\prime = 400 \,\mathrm{MeV}$. \textbf{Right:} Sapphire detector and benchmark DM mass $m_\chi^\prime = 100 \,\mathrm{MeV}$ \textbf{Top:} detector in the Northern Hemisphere at LNGS. \textbf{Bottom:} detector in the Southern Hemisphere at SUPL. For a Germanium detector in the Southern hemisphere (bottom left), we find no closed contours in the $(\sigma_p^\mathrm{SI}, \,\rho_\chi)$ plane.}
    \label{fig:contours}
\end{figure*}

\section{Results}
The projected $p=0.05$ contours in the~$(\sigma_p^{\rm SI},\rho_\chi)$-plane are shown in Fig.~\ref{fig:contours}.
The upper~(lower) panel shows reconstructions for a hypothetical direct detection experiment in the Northern~(Southern) Hemisphere, located at Laboratori Nazionali del Gran Sasso (LNGS, $46^\circ\text{N}$) and Stawell Underground Physics Laboratory (SUPL, $37^\circ\text{S}$) respectively~\cite{Bellotti:1988bw,Urquijo:2016dxd}. In the left~(right) panels, we show results for 400~MeV DM particle and germanium detector (100~MeV DM particle and sapphire detector).

In Fig.~\ref{fig:contours}, we focus on 5 benchmark values for the cross section, $\sigma_p^{\rm SI \prime}$, in the range $10^{-33}$~cm$^2$ - 10$^{-30}$~cm$^2$. While we have performed the analysis over a wider range of cross sections, we do not find closed contours in the ($\sigma^{\rm SI}_p$,~$\rho_\chi$) plane for cross sections smaller than $\sigma_p^{\rm SI \prime}$. This indicates that for our heavier benchmark mass of 400~MeV (left panels), there remains no unconstrained region of parameter space for SI interactions, for which a substantial modulation effect due to Earth-scattering can be observed. We therefore focus in the remainder of this work on the light benchmark of mass 100~MeV (right panels).

We note also that it is also not possible to obtain closed contours in the ($\sigma^{\rm SI}_p$,~$\rho_\chi$) plane when taking only the recoil energies of the observed signal events into account, assuming no knowledge of their timing information. This is because of the strong degeneracy between~$\sigma^{\rm SI}_p$ and~$\rho_\chi$, which can not be broken by the energy data alone.

In contrast, keeping track of the signals' timing and accounting for their modulation signature improves the situation drastically, as seen in the colored contours of Fig.~\ref{fig:contours}, for the scenario of a light DM particle (right panels). 
In the case of the Northern experiment and cross sections above about~$10^{-33}\,\text{cm}^2$, the degeneracy between DM~density and scattering cross section starts to become weaker.
For higher benchmark cross sections, the true density as well as the cross section itself can be reconstructed with increasing precision.
For example, for~$\sigma_p^{\rm SI}=10^{-32}\,\text{cm}^2$ ($\sigma_p^{\rm SI}=10^{-31}\,\text{cm}^2$) we could determine the local DM~density to be~$\rho_\chi = 0.40_{-0.03}^{+0.26} \,\text{GeV cm}^{-3}$ ($\rho_\chi = 0.40_{-0.02}^{+0.01} \,\text{GeV cm}^{-3}$) at 95\%~CL. In these cases, the cross section would be constrained to $\sigma_p^\mathrm{SI} = 1.00_{-0.33}^{+0.07} \times 10^{-32} \,\text{cm}^2$ ($\sigma_p^\mathrm{SI} = 1.00_{-0.01}^{+0.03} \times 10^{-31} \,\text{cm}^2$).

Projected contours for an experiment at SUPL (lower panel) show a similar evolution.
However, the reconstruction of the local DM~density and cross section is generally less precise than in the Northern Hemisphere.
In the case of a benchmark cross section of $\sigma_p^{\rm SI}=10^{-32}\,\text{cm}^2$, a closed contour is obtained though the constraints on $\rho_\chi$ are very wide (extending over the entire range of our analysis, from 0.01 GeV/cm$^{3}$ to 1.0 GeV/cm$^3$). Instead, for a benchmark cross section of $\sigma_p^{\rm SI}=10^{-32}\,\text{cm}^2$, we find $\rho_\chi = 0.40_{-0.19}^{+0.23} \,\text{GeV cm}^{-3}$ and $\sigma_p^\mathrm{SI} = 1.00_{-0.21}^{+3.77} \times 10^{-31} \,\text{cm}^2$.

These results indicate that with future ultra-low threshold detectors, it should be possible to reconstruct both the local DM~density and cross section, if the DM-proton cross section lies within a few orders of magnitude of current constraints, for a DM mass of 100~MeV.

\section{Discussion} 

From these results, the necessity of timing information is obvious. Without time-tagging, it is always possible to reabsorb a change of the local DM~density into a re-scaling of the cross section such that $\rho_\chi \times \sigma_p^{\rm SI}$ remains constant.
The time-dependence of the local DM~distribution in the laboratory, caused by underground scatterings, introduces an additional dependence of the signal on $\sigma_p^{\rm SI}$.
Since this dependence manifests itself through diurnal modulation, knowledge of event timings is the key to disentangling the local DM~density and the cross section.

\begin{figure*}
    \centering
    \includegraphics[width=0.45\textwidth]{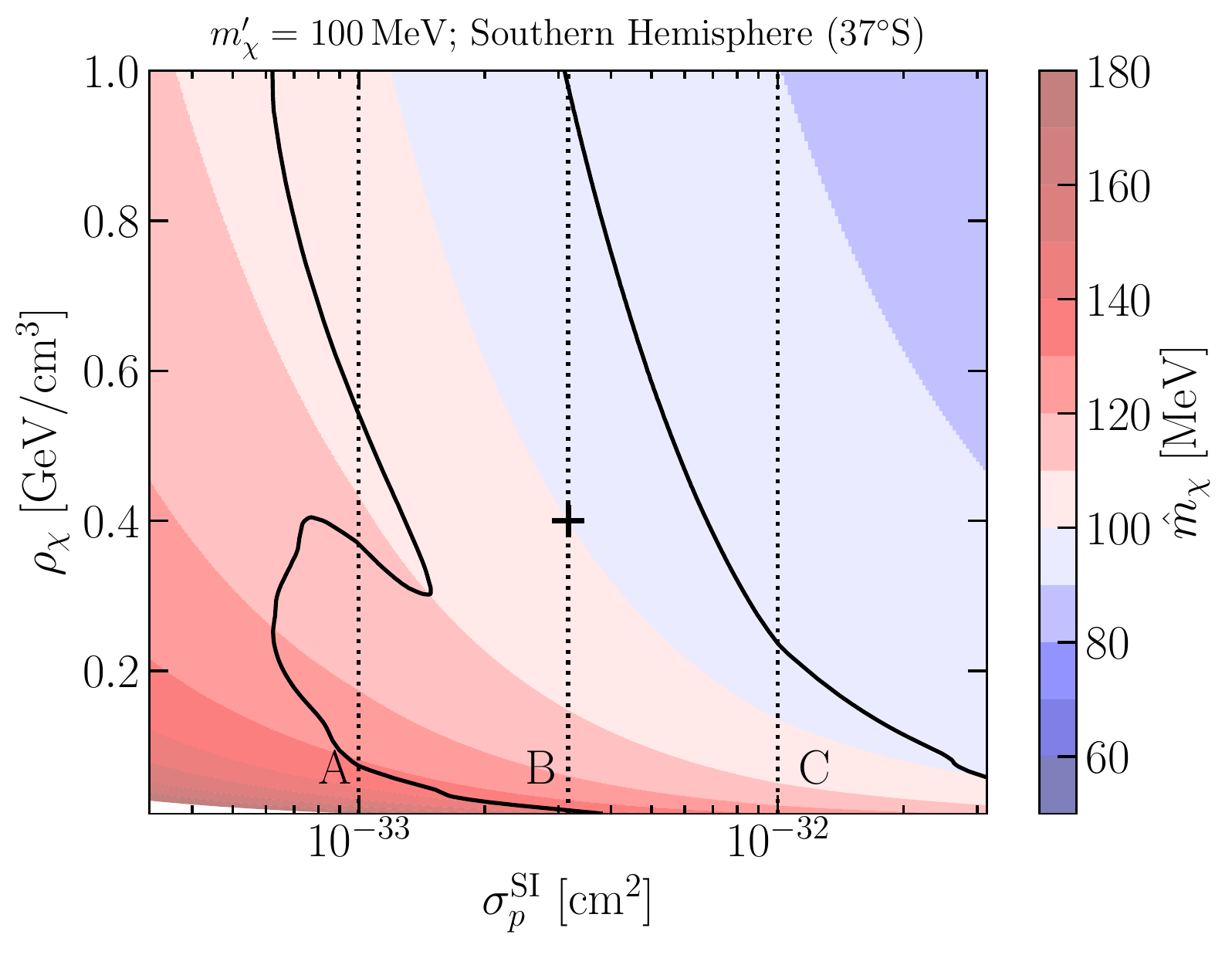}
    \includegraphics[width=0.54\textwidth]{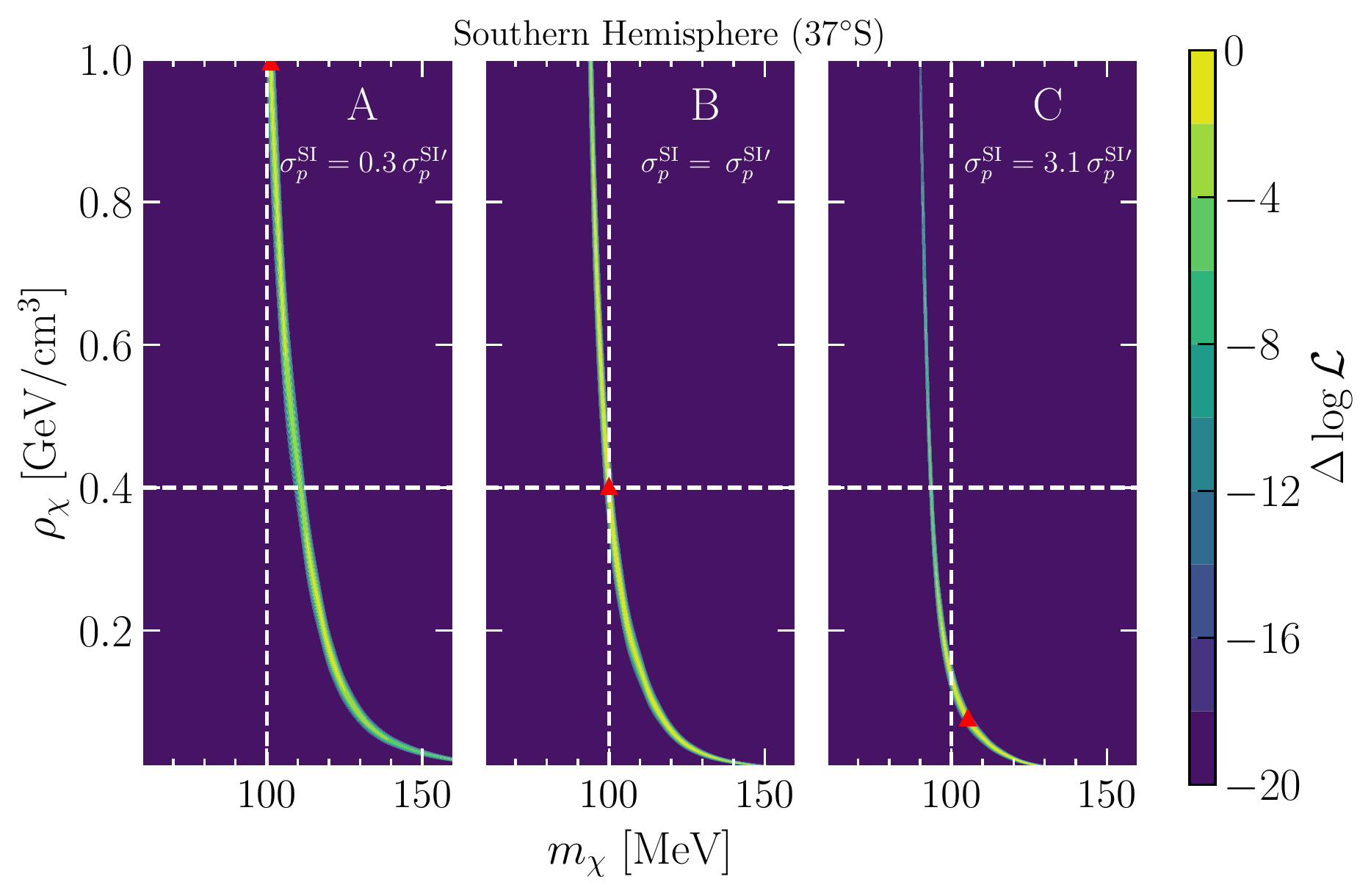}
    \caption{\textbf{Left:} Projected 95\% CL contour for a single benchmark ($\sigma_p^{\rm SI \prime} \approx 3 \times 10^{-32}\,\mathrm{cm}^2, m_\chi' = 100\,\mathrm{MeV}, \rho^\prime_\chi=0.4\,\mathrm{GeV\,cm}^{-3}$, black cross), assuming a sapphire detector in the Southern Hemisphere. The colored shading shows the best fit DM mass at each point.
    \textbf{Right:} Log-likelihood contours in $(m_\chi, \rho_\chi)$ for 3 fixed cross section slices through the parameter space, labelled A--C in the left panel. The white dashed lines show the benchmark values. The log-likelihood is shown relative to the best fit point in each slice (red triangle).}
    \label{fig:Likelihoods}
\end{figure*}

Contrary to our intuition\footnote{Detectors in the Southern Hemisphere are generally more sensitive to these diurnal modulations~\cite{Hasenbalg:1997hs,Emken:2017qmp}.}, we find that experiments in the Northern Hemisphere are generally better suited to measuring the local DM~density.
For an experiment at LNGS, $\Theta(t)$ varies in the range $[4^\circ, 84^\circ]$, so the bulk of the incoming DM~flux reaches the laboratory directly from above at a certain time of day, while it has to pass through a small fraction of Earth's mantle 12 hours later.
Therefore, the experiment switches continuously between being totally exposed to and partially shielded from incoming DM~particles, as illustrated in Fig.~\ref{fig: isodetection angle}. Increasing the cross section increases the modulation amplitude and steadily improves the reconstruction of $\rho_\chi$.
 
However, in order to reach a direct detection experiment in the Southern Hemisphere, most DM~particles need to traverse the planet's bulk mass throughout the day.
An experiment at SUPL is always partially shielded, with $\Theta(t) \in [86^\circ, 167^\circ]$.
For cross sections between~$10^{-34}-10^{-32}\,\text{ cm}^2$, the Earth's stopping power renders the majority of the DM~wind undetectable.  The slower sub-component, which arrives from the opposite direction passing the atmosphere and overburden only, is not yet affected (see again Fig.~\ref{fig: isodetection angle}).
In this regime, the modulation amplitude depends only weakly on the cross section, and estimates of~$\rho_\chi$ and~$\sigma_p^{\rm SI}$ are less precise than in the Northern Hemisphere.
This demonstrates that the determining factor for reconstructing the DM~density is not so much the diurnal modulation's amplitude, but rather its sensitivity to changes in the cross section.

To better understand the contour shapes in Fig.~\ref{fig:contours}, we focus on the benchmark point~$m_\chi^\prime = 100\,\mathrm{MeV}$, $\sigma_p^{\rm SI \prime} \approx 3\times 10^{-33}\,\text{ cm}^2$ and a sapphire experiment in the Southern Hemisphere. 
The left panel of Fig.~\ref{fig:Likelihoods} shows the projected contour and best fit mass at each point in parameter space $(\sigma_p^\mathrm{SI},\rho_\chi)$, while the right panel shows the log-likelihood across three fixed cross section slices (A--C).
In each slice, the curved region where the log-likelihood peaks corresponds roughly to a region where the total number of signal events is constant. The dominant effect is that increasing the DM mass from the benchmark value exponentially increases the number of events above threshold. This is because the typical recoil energy deposited by a 100~MeV particle in a sapphire detector $\mathcal{O}(8\,\mathrm{eV})$, is so close to our assumed threshold of 10~eV. This exponential increase in the number of signal events must be compensated by a decrease in $\rho_\chi$, as seen in the right panel of Fig.~\ref{fig:Likelihoods}.

Focusing on slice~C, overestimating the scattering cross section would mean predicting more events and a greater modulation amplitude.
The former is compensated by a lower best-fit value of $\rho_\chi$, while the latter is compensated by an overestimate of $m_\chi$ (red triangle). Though increasing $m_\chi$ increases the Earth's stopping power\footnote{At fixed cross-section, Earth stopping is more effective for DM closer to nuclear masses, due to the reduced kinematic mismatch~\cite{Kavanagh:2017cru}.}, it also increases the typical recoil energy which can be deposited by a DM particle. For signals close to threshold, this latter effect wins out, meaning that a larger DM mass reduces the overall attenuation of the signal and lowers the modulation amplitude due to Earth scattering. Thus, in slice~C the log-likelihood peaks in a region of higher mass, over a restricted range of values for $\rho_\chi$. In general, the slope of the log-likelihood contours around the peak then determines the uncertainty on $\rho_\chi$ and thereby the contour shape in the left panel of Fig.~\ref{fig:Likelihoods}.

\section{Conclusions}

Provided that DM-matter interactions are sufficiently strong for underground scatterings to occur frequently, signals in direct detection experiments should show a diurnal modulation, which can be exploited to break the degeneracy between~$\rho_\chi$ and~$\sigma_p^{\rm SI}$. We have explored this possibility for a number of benchmarks using MC~simulations. 

For the case of a DM mass of $400 \,\mathrm{MeV}$ (left panel of Fig.~\ref{fig:contours}), we find that both $\rho_\chi$ and~$\sigma_p^{\rm SI}$ can be reconstructed only for cross sections close to $10^{-30}\,\mathrm{cm}^2$, substantially larger than allowed by current constraints. Lighter DM of mass $100 \,\mathrm{MeV}$ (right panel of Fig.~\ref{fig:contours}) should be accessible to near-future low-threshold nuclear recoil searches. In this case, it should be possible to disentangle the local DM density and the scattering cross section for models just below current constraints, in the range $\sigma_p^{\rm SI} \in [10^{-33},\,10^{-30}]\,\mathrm{cm}^2$. For this, detectors in the Northerm Hemisphere offer the best prospects, with a $\mathcal{O}(50\%)$ ($\mathcal{O}(5\%)$) reconstruction of $\rho_\chi$ possible for $\sigma_p^{\rm SI}=10^{-32}\,\text{cm}^2$ ($\sigma_p^{\rm SI}=10^{-31}\,\text{cm}^2$).

This is the first demonstration that it is possible to measure the local Dark Matter density directly in the laboratory and further motivates the search for light, strongly-interacting DM with low-threshold detectors.

\acknowledgments

RC and TE were supported by the Knut and Alice Wallenberg Foundation (PI, Jan Conrad).~RC also acknowledges support from an individual research grant from the Swedish Research Council, dnr. 2018-05029. BJK would like to thank the Spanish Agencia Estatal de Investigaci\'on (AEI, MICIU) for the support to the Unidad de Excelencia Mar\'ia de Maeztu Instituto de F\'isica de Cantabria, ref. MDM-2017-0765. Some of the computations were performed on resources provided by the Swedish National Infrastructure for Computing (SNIC) at NSC, as well as the Dutch national e-infrastructure with the support of SURF Cooperative.

We would like to thank the Munich Institute for Astro- and Particle Physics (MIAPP) where part of this work
was developed. Finally, we acknowledge the use of the Python scientific computing packages NumPy \cite{numpy} and SciPy \cite{scipy}, as well as the graphics environment Matplotlib \cite{Hunter:2007}.

\bibliography{references}

\begin{thebibliography}{123}%
\makeatletter
\providecommand \@ifxundefined [1]{%
 \@ifx{#1\undefined}
}%
\providecommand \@ifnum [1]{%
 \ifnum #1\expandafter \@firstoftwo
 \else \expandafter \@secondoftwo
 \fi
}%
\providecommand \@ifx [1]{%
 \ifx #1\expandafter \@firstoftwo
 \else \expandafter \@secondoftwo
 \fi
}%
\providecommand \natexlab [1]{#1}%
\providecommand \enquote  [1]{``#1''}%
\providecommand \bibnamefont  [1]{#1}%
\providecommand \bibfnamefont [1]{#1}%
\providecommand \citenamefont [1]{#1}%
\providecommand \href@noop [0]{\@secondoftwo}%
\providecommand \href [0]{\begingroup \@sanitize@url \@href}%
\providecommand \@href[1]{\@@startlink{#1}\@@href}%
\providecommand \@@href[1]{\endgroup#1\@@endlink}%
\providecommand \@sanitize@url [0]{\catcode `\\12\catcode `\$12\catcode
  `\&12\catcode `\#12\catcode `\^12\catcode `\_12\catcode `\%12\relax}%
\providecommand \@@startlink[1]{}%
\providecommand \@@endlink[0]{}%
\providecommand \url  [0]{\begingroup\@sanitize@url \@url }%
\providecommand \@url [1]{\endgroup\@href {#1}{\urlprefix }}%
\providecommand \urlprefix  [0]{URL }%
\providecommand \Eprint [0]{\href }%
\providecommand \doibase [0]{http://dx.doi.org/}%
\providecommand \selectlanguage [0]{\@gobble}%
\providecommand \bibinfo  [0]{\@secondoftwo}%
\providecommand \bibfield  [0]{\@secondoftwo}%
\providecommand \translation [1]{[#1]}%
\providecommand \BibitemOpen [0]{}%
\providecommand \bibitemStop [0]{}%
\providecommand \bibitemNoStop [0]{.\EOS\space}%
\providecommand \EOS [0]{\spacefactor3000\relax}%
\providecommand \BibitemShut  [1]{\csname bibitem#1\endcsname}%
\let\auto@bib@innerbib\@empty
\bibitem [{\citenamefont {Bertone}\ \emph {et~al.}(2005)\citenamefont
  {Bertone}, \citenamefont {Hooper},\ and\ \citenamefont
  {Silk}}]{Bertone:2004pz}%
  \BibitemOpen
  \bibfield  {author} {\bibinfo {author} {\bibfnamefont {G.}~\bibnamefont
  {Bertone}}, \bibinfo {author} {\bibfnamefont {D.}~\bibnamefont {Hooper}}, \
  and\ \bibinfo {author} {\bibfnamefont {J.}~\bibnamefont {Silk}},\ }\href
  {\doibase 10.1016/j.physrep.2004.08.031} {\bibfield  {journal} {\bibinfo
  {journal} {Phys. Rept.}\ }\textbf {\bibinfo {volume} {405}},\ \bibinfo
  {pages} {279} (\bibinfo {year} {2005})},\ \Eprint
  {http://arxiv.org/abs/hep-ph/0404175} {arXiv:hep-ph/0404175 [hep-ph]}
  \BibitemShut {NoStop}%
\bibitem [{\citenamefont {Bertone}\ and\ \citenamefont
  {Hooper}(2018)}]{Bertone:2016nfn}%
  \BibitemOpen
  \bibfield  {author} {\bibinfo {author} {\bibfnamefont {G.}~\bibnamefont
  {Bertone}}\ and\ \bibinfo {author} {\bibfnamefont {D.}~\bibnamefont
  {Hooper}},\ }\href {\doibase 10.1103/RevModPhys.90.045002} {\bibfield
  {journal} {\bibinfo  {journal} {Rev. Mod. Phys.}\ }\textbf {\bibinfo {volume}
  {90}},\ \bibinfo {pages} {045002} (\bibinfo {year} {2018})},\ \Eprint
  {http://arxiv.org/abs/1605.04909} {arXiv:1605.04909 [astro-ph.CO]}
  \BibitemShut {NoStop}%
\bibitem [{\citenamefont {{\L}okas}(2002)}]{Lokas2002Jul}%
  \BibitemOpen
  \bibfield  {author} {\bibinfo {author} {\bibfnamefont {E.~L.}\ \bibnamefont
  {{\L}okas}},\ }\href {\doibase 10.1046/j.1365-8711.2002.05457.x} {\bibfield
  {journal} {\bibinfo  {journal} {Mon. Not. R. Astron. Soc.}\ }\textbf
  {\bibinfo {volume} {333}},\ \bibinfo {pages} {697} (\bibinfo {year}
  {2002})}\BibitemShut {NoStop}%
\bibitem [{\citenamefont {Ade}\ \emph {et~al.}(2016)\citenamefont {Ade} \emph
  {et~al.}}]{Ade:2015xua}%
  \BibitemOpen
  \bibfield  {author} {\bibinfo {author} {\bibfnamefont {P.~A.~R.}\
  \bibnamefont {Ade}} \emph {et~al.} (\bibinfo {collaboration} {Planck}),\
  }\href {\doibase 10.1051/0004-6361/201525830} {\bibfield  {journal} {\bibinfo
   {journal} {Astron. Astrophys.}\ }\textbf {\bibinfo {volume} {594}},\
  \bibinfo {pages} {A13} (\bibinfo {year} {2016})},\ \Eprint
  {http://arxiv.org/abs/1502.01589} {arXiv:1502.01589 [astro-ph.CO]}
  \BibitemShut {NoStop}%
\bibitem [{\citenamefont {Brown}\ \emph {et~al.}(2018)\citenamefont {Brown}
  \emph {et~al.}}]{Brown:2018dum}%
  \BibitemOpen
  \bibfield  {author} {\bibinfo {author} {\bibfnamefont {A.~G.~A.}\
  \bibnamefont {Brown}} \emph {et~al.} (\bibinfo {collaboration} {Gaia}),\
  }\href {\doibase 10.1051/0004-6361/201833051} {\bibfield  {journal} {\bibinfo
   {journal} {Astron. Astrophys.}\ }\textbf {\bibinfo {volume} {616}},\
  \bibinfo {pages} {A1} (\bibinfo {year} {2018})},\ \Eprint
  {http://arxiv.org/abs/1804.09365} {arXiv:1804.09365 [astro-ph.GA]}
  \BibitemShut {NoStop}%
\bibitem [{\citenamefont {Moniez}(2010)}]{Moniez:2010zt}%
  \BibitemOpen
  \bibfield  {author} {\bibinfo {author} {\bibfnamefont {M.}~\bibnamefont
  {Moniez}},\ }\href {\doibase 10.1007/s10714-009-0925-4} {\bibfield  {journal}
  {\bibinfo  {journal} {Gen. Rel. Grav.}\ }\textbf {\bibinfo {volume} {42}},\
  \bibinfo {pages} {2047} (\bibinfo {year} {2010})},\ \Eprint
  {http://arxiv.org/abs/1001.2707} {arXiv:1001.2707 [astro-ph.GA]} \BibitemShut
  {NoStop}%
\bibitem [{\citenamefont {Callingham}\ \emph {et~al.}(2018)\citenamefont
  {Callingham}, \citenamefont {Cautun}, \citenamefont {Deason}, \citenamefont
  {Frenk}, \citenamefont {Wang}, \citenamefont {Gómez}, \citenamefont {Grand},
  \citenamefont {Marinacci},\ and\ \citenamefont
  {Pakmor}}]{Callingham:2018vcf}%
  \BibitemOpen
  \bibfield  {author} {\bibinfo {author} {\bibfnamefont {T.}~\bibnamefont
  {Callingham}}, \bibinfo {author} {\bibfnamefont {M.}~\bibnamefont {Cautun}},
  \bibinfo {author} {\bibfnamefont {A.~J.}\ \bibnamefont {Deason}}, \bibinfo
  {author} {\bibfnamefont {C.~S.}\ \bibnamefont {Frenk}}, \bibinfo {author}
  {\bibfnamefont {W.}~\bibnamefont {Wang}}, \bibinfo {author} {\bibfnamefont
  {F.~A.}\ \bibnamefont {Gómez}}, \bibinfo {author} {\bibfnamefont {R.~J.~J.}\
  \bibnamefont {Grand}}, \bibinfo {author} {\bibfnamefont {F.}~\bibnamefont
  {Marinacci}}, \ and\ \bibinfo {author} {\bibfnamefont {R.}~\bibnamefont
  {Pakmor}},\ }\href {\doibase 10.1093/mnras/stz365} {\  (\bibinfo {year}
  {2018}),\ 10.1093/mnras/stz365},\ \Eprint {http://arxiv.org/abs/1808.10456}
  {arXiv:1808.10456 [astro-ph.GA]} \BibitemShut {NoStop}%
\bibitem [{\citenamefont {Tremaine}(1990)}]{Tremaine1990}%
  \BibitemOpen
  \bibfield  {author} {\bibinfo {author} {\bibfnamefont {S.}~\bibnamefont
  {Tremaine}},\ }\href {\doibase 10.1007/978-94-009-0565-8_3} {\bibfield
  {journal} {\bibinfo  {journal} {SpringerLink}\ ,\ \bibinfo {pages} {37}}
  (\bibinfo {year} {1990})}\BibitemShut {NoStop}%
\bibitem [{\citenamefont {Kuijken}\ and\ \citenamefont
  {Gilmore}(1989)}]{Kuijken:1989hu}%
  \BibitemOpen
  \bibfield  {author} {\bibinfo {author} {\bibfnamefont {K.}~\bibnamefont
  {Kuijken}}\ and\ \bibinfo {author} {\bibfnamefont {G.}~\bibnamefont
  {Gilmore}},\ }\href@noop {} {\bibfield  {journal} {\bibinfo  {journal} {Mon.
  Not. Roy. Astron. Soc.}\ }\textbf {\bibinfo {volume} {239}},\ \bibinfo
  {pages} {605} (\bibinfo {year} {1989})}\BibitemShut {NoStop}%
\bibitem [{\citenamefont {{Moni Bidin}}\ \emph {et~al.}(2012)\citenamefont
  {{Moni Bidin}}, \citenamefont {{Carraro}}, \citenamefont {{M{\'e}ndez}},\
  and\ \citenamefont {{Smith}}}]{MoniBidin2012}%
  \BibitemOpen
  \bibfield  {author} {\bibinfo {author} {\bibfnamefont {C.}~\bibnamefont
  {{Moni Bidin}}}, \bibinfo {author} {\bibfnamefont {G.}~\bibnamefont
  {{Carraro}}}, \bibinfo {author} {\bibfnamefont {R.~A.}\ \bibnamefont
  {{M{\'e}ndez}}}, \ and\ \bibinfo {author} {\bibfnamefont {R.}~\bibnamefont
  {{Smith}}},\ }\href {\doibase 10.1088/0004-637X/751/1/30} {\bibfield
  {journal} {\bibinfo  {journal} {Astrophys. J.}\ }\textbf {\bibinfo {volume}
  {751}},\ \bibinfo {eid} {30} (\bibinfo {year} {2012})},\ \Eprint
  {http://arxiv.org/abs/1204.3924} {arXiv:1204.3924} \BibitemShut {NoStop}%
\bibitem [{\citenamefont {Bovy}\ and\ \citenamefont
  {Tremaine}(2012)}]{Bovy:2012tw}%
  \BibitemOpen
  \bibfield  {author} {\bibinfo {author} {\bibfnamefont {J.}~\bibnamefont
  {Bovy}}\ and\ \bibinfo {author} {\bibfnamefont {S.}~\bibnamefont
  {Tremaine}},\ }\href {\doibase 10.1088/0004-637X/756/1/89} {\bibfield
  {journal} {\bibinfo  {journal} {Astrophys. J.}\ }\textbf {\bibinfo {volume}
  {756}},\ \bibinfo {pages} {89} (\bibinfo {year} {2012})},\ \Eprint
  {http://arxiv.org/abs/1205.4033} {arXiv:1205.4033 [astro-ph.GA]} \BibitemShut
  {NoStop}%
\bibitem [{\citenamefont {{Garbari}}\ \emph {et~al.}(2012)\citenamefont
  {{Garbari}}, \citenamefont {{Liu}}, \citenamefont {{Read}},\ and\
  \citenamefont {{Lake}}}]{Garbari2012}%
  \BibitemOpen
  \bibfield  {author} {\bibinfo {author} {\bibfnamefont {S.}~\bibnamefont
  {{Garbari}}}, \bibinfo {author} {\bibfnamefont {C.}~\bibnamefont {{Liu}}},
  \bibinfo {author} {\bibfnamefont {J.~I.}\ \bibnamefont {{Read}}}, \ and\
  \bibinfo {author} {\bibfnamefont {G.}~\bibnamefont {{Lake}}},\ }\href
  {\doibase 10.1111/j.1365-2966.2012.21608.x} {\bibfield  {journal} {\bibinfo
  {journal} {MNRAS}\ }\textbf {\bibinfo {volume} {425}},\ \bibinfo {pages}
  {1445} (\bibinfo {year} {2012})},\ \Eprint {http://arxiv.org/abs/1206.0015}
  {arXiv:1206.0015 [astro-ph.GA]} \BibitemShut {NoStop}%
\bibitem [{\citenamefont {{Smith}}\ \emph {et~al.}(2012)\citenamefont
  {{Smith}}, \citenamefont {{Whiteoak}},\ and\ \citenamefont
  {{Evans}}}]{Smith2012}%
  \BibitemOpen
  \bibfield  {author} {\bibinfo {author} {\bibfnamefont {M.~C.}\ \bibnamefont
  {{Smith}}}, \bibinfo {author} {\bibfnamefont {S.~H.}\ \bibnamefont
  {{Whiteoak}}}, \ and\ \bibinfo {author} {\bibfnamefont {N.~W.}\ \bibnamefont
  {{Evans}}},\ }\href {\doibase 10.1088/0004-637X/746/2/181} {\bibfield
  {journal} {\bibinfo  {journal} {Astrophys. J.}\ }\textbf {\bibinfo {volume}
  {746}},\ \bibinfo {eid} {181} (\bibinfo {year} {2012})},\ \Eprint
  {http://arxiv.org/abs/1111.6920} {arXiv:1111.6920 [astro-ph.GA]} \BibitemShut
  {NoStop}%
\bibitem [{\citenamefont {{Zhang}}\ \emph {et~al.}(2013)\citenamefont
  {{Zhang}}, \citenamefont {{Rix}}, \citenamefont {{van de Ven}}, \citenamefont
  {{Bovy}}, \citenamefont {{Liu}},\ and\ \citenamefont {{Zhao}}}]{Zhang2013}%
  \BibitemOpen
  \bibfield  {author} {\bibinfo {author} {\bibfnamefont {L.}~\bibnamefont
  {{Zhang}}}, \bibinfo {author} {\bibfnamefont {H.-W.}\ \bibnamefont {{Rix}}},
  \bibinfo {author} {\bibfnamefont {G.}~\bibnamefont {{van de Ven}}}, \bibinfo
  {author} {\bibfnamefont {J.}~\bibnamefont {{Bovy}}}, \bibinfo {author}
  {\bibfnamefont {C.}~\bibnamefont {{Liu}}}, \ and\ \bibinfo {author}
  {\bibfnamefont {G.}~\bibnamefont {{Zhao}}},\ }\href {\doibase
  10.1088/0004-637X/772/2/108} {\bibfield  {journal} {\bibinfo  {journal}
  {Astrophys. J.}\ }\textbf {\bibinfo {volume} {772}},\ \bibinfo {eid} {108}
  (\bibinfo {year} {2013})},\ \Eprint {http://arxiv.org/abs/1209.0256}
  {arXiv:1209.0256} \BibitemShut {NoStop}%
\bibitem [{\citenamefont {Bovy}\ and\ \citenamefont
  {Rix}(2013)}]{Bovy:2013raa}%
  \BibitemOpen
  \bibfield  {author} {\bibinfo {author} {\bibfnamefont {J.}~\bibnamefont
  {Bovy}}\ and\ \bibinfo {author} {\bibfnamefont {H.-W.}\ \bibnamefont {Rix}},\
  }\href {\doibase 10.1088/0004-637X/779/2/115} {\bibfield  {journal} {\bibinfo
   {journal} {Astrophys. J.}\ }\textbf {\bibinfo {volume} {779}},\ \bibinfo
  {pages} {115} (\bibinfo {year} {2013})},\ \Eprint
  {http://arxiv.org/abs/1309.0809} {arXiv:1309.0809 [astro-ph.GA]} \BibitemShut
  {NoStop}%
\bibitem [{\citenamefont {Bienaymé}\ \emph {et~al.}(2014)\citenamefont
  {Bienaymé} \emph {et~al.}}]{Bienayme:2014kva}%
  \BibitemOpen
  \bibfield  {author} {\bibinfo {author} {\bibfnamefont {O.}~\bibnamefont
  {Bienaymé}} \emph {et~al.},\ }\href {\doibase 10.1051/0004-6361/201424478}
  {\bibfield  {journal} {\bibinfo  {journal} {Astron. Astrophys.}\ }\textbf
  {\bibinfo {volume} {571}},\ \bibinfo {pages} {A92} (\bibinfo {year}
  {2014})},\ \Eprint {http://arxiv.org/abs/1406.6896} {arXiv:1406.6896
  [astro-ph.GA]} \BibitemShut {NoStop}%
\bibitem [{\citenamefont {McKee}\ \emph {et~al.}(2015)\citenamefont {McKee},
  \citenamefont {Parravano},\ and\ \citenamefont {Hollenbach}}]{McKee:2015hwa}%
  \BibitemOpen
  \bibfield  {author} {\bibinfo {author} {\bibfnamefont {C.~F.}\ \bibnamefont
  {McKee}}, \bibinfo {author} {\bibfnamefont {A.}~\bibnamefont {Parravano}}, \
  and\ \bibinfo {author} {\bibfnamefont {D.~J.}\ \bibnamefont {Hollenbach}},\
  }\href {\doibase 10.1088/0004-637X, 10.1088/0004-637X/814/1/13} {\bibfield
  {journal} {\bibinfo  {journal} {Astrophys. J.}\ }\textbf {\bibinfo {volume}
  {814}},\ \bibinfo {pages} {13} (\bibinfo {year} {2015})},\ \Eprint
  {http://arxiv.org/abs/1509.05334} {arXiv:1509.05334 [astro-ph.GA]}
  \BibitemShut {NoStop}%
\bibitem [{\citenamefont {Xia}\ \emph {et~al.}(2016)\citenamefont {Xia},
  \citenamefont {Liu}, \citenamefont {Mao}, \citenamefont {Song}, \citenamefont
  {Zhang}, \citenamefont {Long}, \citenamefont {Zhang}, \citenamefont {Hou},
  \citenamefont {Wang},\ and\ \citenamefont {Wu}}]{Xia:2015agz}%
  \BibitemOpen
  \bibfield  {author} {\bibinfo {author} {\bibfnamefont {Q.}~\bibnamefont
  {Xia}}, \bibinfo {author} {\bibfnamefont {C.}~\bibnamefont {Liu}}, \bibinfo
  {author} {\bibfnamefont {S.}~\bibnamefont {Mao}}, \bibinfo {author}
  {\bibfnamefont {Y.}~\bibnamefont {Song}}, \bibinfo {author} {\bibfnamefont
  {L.}~\bibnamefont {Zhang}}, \bibinfo {author} {\bibfnamefont {R.~J.}\
  \bibnamefont {Long}}, \bibinfo {author} {\bibfnamefont {Y.}~\bibnamefont
  {Zhang}}, \bibinfo {author} {\bibfnamefont {Y.}~\bibnamefont {Hou}}, \bibinfo
  {author} {\bibfnamefont {Y.}~\bibnamefont {Wang}}, \ and\ \bibinfo {author}
  {\bibfnamefont {Y.}~\bibnamefont {Wu}},\ }\href {\doibase
  10.1093/mnras/stw565} {\bibfield  {journal} {\bibinfo  {journal} {Mon. Not.
  Roy. Astron. Soc.}\ }\textbf {\bibinfo {volume} {458}},\ \bibinfo {pages}
  {3839} (\bibinfo {year} {2016})},\ \Eprint {http://arxiv.org/abs/1510.06810}
  {arXiv:1510.06810 [astro-ph.GA]} \BibitemShut {NoStop}%
\bibitem [{\citenamefont {Sivertsson}\ \emph {et~al.}(2018)\citenamefont
  {Sivertsson}, \citenamefont {Silverwood}, \citenamefont {Read}, \citenamefont
  {Bertone},\ and\ \citenamefont {Steger}}]{Sivertsson:2017rkp}%
  \BibitemOpen
  \bibfield  {author} {\bibinfo {author} {\bibfnamefont {S.}~\bibnamefont
  {Sivertsson}}, \bibinfo {author} {\bibfnamefont {H.}~\bibnamefont
  {Silverwood}}, \bibinfo {author} {\bibfnamefont {J.~I.}\ \bibnamefont
  {Read}}, \bibinfo {author} {\bibfnamefont {G.}~\bibnamefont {Bertone}}, \
  and\ \bibinfo {author} {\bibfnamefont {P.}~\bibnamefont {Steger}},\ }\href
  {\doibase 10.1093/mnras/sty977} {\bibfield  {journal} {\bibinfo  {journal}
  {Mon. Not. Roy. Astron. Soc.}\ }\textbf {\bibinfo {volume} {478}},\ \bibinfo
  {pages} {1677} (\bibinfo {year} {2018})},\ \Eprint
  {http://arxiv.org/abs/1708.07836} {arXiv:1708.07836 [astro-ph.GA]}
  \BibitemShut {NoStop}%
\bibitem [{\citenamefont {Buch}\ \emph {et~al.}(2019)\citenamefont {Buch},
  \citenamefont {Leung},\ and\ \citenamefont {Fan}}]{Buch:2018qdr}%
  \BibitemOpen
  \bibfield  {author} {\bibinfo {author} {\bibfnamefont {J.}~\bibnamefont
  {Buch}}, \bibinfo {author} {\bibfnamefont {S.~C.~J.}\ \bibnamefont {Leung}},
  \ and\ \bibinfo {author} {\bibfnamefont {J.}~\bibnamefont {Fan}},\ }\href
  {\doibase 10.1088/1475-7516/2019/04/026} {\bibfield  {journal} {\bibinfo
  {journal} {JCAP}\ }\textbf {\bibinfo {volume} {1904}},\ \bibinfo {pages}
  {026} (\bibinfo {year} {2019})},\ \Eprint {http://arxiv.org/abs/1808.05603}
  {arXiv:1808.05603 [astro-ph.GA]} \BibitemShut {NoStop}%
\bibitem [{\citenamefont {Salucci}\ \emph {et~al.}(2010)\citenamefont
  {Salucci}, \citenamefont {Nesti}, \citenamefont {Gentile},\ and\
  \citenamefont {Martins}}]{Salucci:2010qr}%
  \BibitemOpen
  \bibfield  {author} {\bibinfo {author} {\bibfnamefont {P.}~\bibnamefont
  {Salucci}}, \bibinfo {author} {\bibfnamefont {F.}~\bibnamefont {Nesti}},
  \bibinfo {author} {\bibfnamefont {G.}~\bibnamefont {Gentile}}, \ and\
  \bibinfo {author} {\bibfnamefont {C.~F.}\ \bibnamefont {Martins}},\ }\href
  {\doibase 10.1051/0004-6361/201014385} {\bibfield  {journal} {\bibinfo
  {journal} {Astron. Astrophys.}\ }\textbf {\bibinfo {volume} {523}},\ \bibinfo
  {pages} {A83} (\bibinfo {year} {2010})},\ \Eprint
  {http://arxiv.org/abs/1003.3101} {arXiv:1003.3101 [astro-ph.GA]} \BibitemShut
  {NoStop}%
\bibitem [{\citenamefont {Catena}\ and\ \citenamefont
  {Ullio}(2010)}]{Catena:2009mf}%
  \BibitemOpen
  \bibfield  {author} {\bibinfo {author} {\bibfnamefont {R.}~\bibnamefont
  {Catena}}\ and\ \bibinfo {author} {\bibfnamefont {P.}~\bibnamefont {Ullio}},\
  }\href {\doibase 10.1088/1475-7516/2010/08/004} {\bibfield  {journal}
  {\bibinfo  {journal} {JCAP}\ }\textbf {\bibinfo {volume} {1008}},\ \bibinfo
  {pages} {004} (\bibinfo {year} {2010})},\ \Eprint
  {http://arxiv.org/abs/0907.0018} {arXiv:0907.0018 [astro-ph.CO]} \BibitemShut
  {NoStop}%
\bibitem [{\citenamefont {{Weber}}\ and\ \citenamefont {{de
  Boer}}(2010)}]{Weber2010}%
  \BibitemOpen
  \bibfield  {author} {\bibinfo {author} {\bibfnamefont {M.}~\bibnamefont
  {{Weber}}}\ and\ \bibinfo {author} {\bibfnamefont {W.}~\bibnamefont {{de
  Boer}}},\ }\href {\doibase 10.1051/0004-6361/200913381} {\bibfield  {journal}
  {\bibinfo  {journal} {Astron. Astrophys.}\ }\textbf {\bibinfo {volume}
  {509}},\ \bibinfo {eid} {A25} (\bibinfo {year} {2010})},\ \Eprint
  {http://arxiv.org/abs/0910.4272} {arXiv:0910.4272 [astro-ph.CO]} \BibitemShut
  {NoStop}%
\bibitem [{\citenamefont {Iocco}\ \emph {et~al.}(2011)\citenamefont {Iocco},
  \citenamefont {Pato}, \citenamefont {Bertone},\ and\ \citenamefont
  {Jetzer}}]{Iocco:2011jz}%
  \BibitemOpen
  \bibfield  {author} {\bibinfo {author} {\bibfnamefont {F.}~\bibnamefont
  {Iocco}}, \bibinfo {author} {\bibfnamefont {M.}~\bibnamefont {Pato}},
  \bibinfo {author} {\bibfnamefont {G.}~\bibnamefont {Bertone}}, \ and\
  \bibinfo {author} {\bibfnamefont {P.}~\bibnamefont {Jetzer}},\ }\href
  {\doibase 10.1088/1475-7516/2011/11/029} {\bibfield  {journal} {\bibinfo
  {journal} {JCAP}\ }\textbf {\bibinfo {volume} {1111}},\ \bibinfo {pages}
  {029} (\bibinfo {year} {2011})},\ \Eprint {http://arxiv.org/abs/1107.5810}
  {arXiv:1107.5810 [astro-ph.GA]} \BibitemShut {NoStop}%
\bibitem [{\citenamefont {{McMillan}}\ and\ \citenamefont
  {{Binney}}(2010)}]{McMillan2010}%
  \BibitemOpen
  \bibfield  {author} {\bibinfo {author} {\bibfnamefont {P.~J.}\ \bibnamefont
  {{McMillan}}}\ and\ \bibinfo {author} {\bibfnamefont {J.~J.}\ \bibnamefont
  {{Binney}}},\ }\href {\doibase 10.1111/j.1365-2966.2009.15932.x} {\bibfield
  {journal} {\bibinfo  {journal} {MNRAS}\ }\textbf {\bibinfo {volume} {402}},\
  \bibinfo {pages} {934} (\bibinfo {year} {2010})},\ \Eprint
  {http://arxiv.org/abs/0907.4685} {arXiv:0907.4685} \BibitemShut {NoStop}%
\bibitem [{\citenamefont {Pato}\ \emph {et~al.}(2015)\citenamefont {Pato},
  \citenamefont {Iocco},\ and\ \citenamefont {Bertone}}]{Pato:2015dua}%
  \BibitemOpen
  \bibfield  {author} {\bibinfo {author} {\bibfnamefont {M.}~\bibnamefont
  {Pato}}, \bibinfo {author} {\bibfnamefont {F.}~\bibnamefont {Iocco}}, \ and\
  \bibinfo {author} {\bibfnamefont {G.}~\bibnamefont {Bertone}},\ }\href
  {\doibase 10.1088/1475-7516/2015/12/001} {\bibfield  {journal} {\bibinfo
  {journal} {JCAP}\ }\textbf {\bibinfo {volume} {1512}},\ \bibinfo {pages}
  {001} (\bibinfo {year} {2015})},\ \Eprint {http://arxiv.org/abs/1504.06324}
  {arXiv:1504.06324 [astro-ph.GA]} \BibitemShut {NoStop}%
\bibitem [{\citenamefont {{Huang}}\ \emph {et~al.}(2016)\citenamefont
  {{Huang}}, \citenamefont {{Liu}}, \citenamefont {{Yuan}}, \citenamefont
  {{Xiang}}, \citenamefont {{Zhang}}, \citenamefont {{Chen}}, \citenamefont
  {{Ren}}, \citenamefont {{Wang}}, \citenamefont {{Zhang}}, \citenamefont
  {{Hou}}, \citenamefont {{Wang}},\ and\ \citenamefont {{Cao}}}]{Huang2016}%
  \BibitemOpen
  \bibfield  {author} {\bibinfo {author} {\bibfnamefont {Y.}~\bibnamefont
  {{Huang}}}, \bibinfo {author} {\bibfnamefont {X.-W.}\ \bibnamefont {{Liu}}},
  \bibinfo {author} {\bibfnamefont {H.-B.}\ \bibnamefont {{Yuan}}}, \bibinfo
  {author} {\bibfnamefont {M.-S.}\ \bibnamefont {{Xiang}}}, \bibinfo {author}
  {\bibfnamefont {H.-W.}\ \bibnamefont {{Zhang}}}, \bibinfo {author}
  {\bibfnamefont {B.-Q.}\ \bibnamefont {{Chen}}}, \bibinfo {author}
  {\bibfnamefont {J.-J.}\ \bibnamefont {{Ren}}}, \bibinfo {author}
  {\bibfnamefont {C.}~\bibnamefont {{Wang}}}, \bibinfo {author} {\bibfnamefont
  {Y.}~\bibnamefont {{Zhang}}}, \bibinfo {author} {\bibfnamefont {Y.-H.}\
  \bibnamefont {{Hou}}}, \bibinfo {author} {\bibfnamefont {Y.-F.}\ \bibnamefont
  {{Wang}}}, \ and\ \bibinfo {author} {\bibfnamefont {Z.-H.}\ \bibnamefont
  {{Cao}}},\ }\href {\doibase 10.1093/mnras/stw2096} {\bibfield  {journal}
  {\bibinfo  {journal} {MNRAS}\ }\textbf {\bibinfo {volume} {463}},\ \bibinfo
  {pages} {2623} (\bibinfo {year} {2016})},\ \Eprint
  {http://arxiv.org/abs/1604.01216} {arXiv:1604.01216} \BibitemShut {NoStop}%
\bibitem [{\citenamefont {{McMillan}}(2017)}]{McMillan2017}%
  \BibitemOpen
  \bibfield  {author} {\bibinfo {author} {\bibfnamefont {P.~J.}\ \bibnamefont
  {{McMillan}}},\ }\href {\doibase 10.1093/mnras/stw2759} {\bibfield  {journal}
  {\bibinfo  {journal} {MNRAS}\ }\textbf {\bibinfo {volume} {465}},\ \bibinfo
  {pages} {76} (\bibinfo {year} {2017})},\ \Eprint
  {http://arxiv.org/abs/1608.00971} {arXiv:1608.00971} \BibitemShut {NoStop}%
\bibitem [{\citenamefont {{de Salas}}\ \emph {et~al.}(2019)\citenamefont {{de
  Salas}}, \citenamefont {{Malhan}}, \citenamefont {{Freese}}, \citenamefont
  {{Hattori}},\ and\ \citenamefont {{Valluri}}}]{2019JCAP...10..037D}%
  \BibitemOpen
  \bibfield  {author} {\bibinfo {author} {\bibfnamefont {P.~F.}\ \bibnamefont
  {{de Salas}}}, \bibinfo {author} {\bibfnamefont {K.}~\bibnamefont
  {{Malhan}}}, \bibinfo {author} {\bibfnamefont {K.}~\bibnamefont {{Freese}}},
  \bibinfo {author} {\bibfnamefont {K.}~\bibnamefont {{Hattori}}}, \ and\
  \bibinfo {author} {\bibfnamefont {M.}~\bibnamefont {{Valluri}}},\ }\href
  {\doibase 10.1088/1475-7516/2019/10/037} {\bibfield  {journal} {\bibinfo
  {journal} {"JCAP"}\ }\textbf {\bibinfo {volume} {2019}},\ \bibinfo {eid}
  {037} (\bibinfo {year} {2019})},\ \Eprint {http://arxiv.org/abs/1906.06133}
  {arXiv:1906.06133 [astro-ph.GA]} \BibitemShut {NoStop}%
\bibitem [{\citenamefont {Peter}(2011)}]{Peter:2011eu}%
  \BibitemOpen
  \bibfield  {author} {\bibinfo {author} {\bibfnamefont {A.~H.~G.}\
  \bibnamefont {Peter}},\ }\href {\doibase 10.1103/PhysRevD.83.125029}
  {\bibfield  {journal} {\bibinfo  {journal} {Phys. Rev.}\ }\textbf {\bibinfo
  {volume} {D83}},\ \bibinfo {pages} {125029} (\bibinfo {year} {2011})},\
  \Eprint {http://arxiv.org/abs/1103.5145} {arXiv:1103.5145 [astro-ph.CO]}
  \BibitemShut {NoStop}%
\bibitem [{\citenamefont {Fairbairn}\ \emph {et~al.}(2013)\citenamefont
  {Fairbairn}, \citenamefont {Douce},\ and\ \citenamefont
  {Swift}}]{Fairbairn:2012zs}%
  \BibitemOpen
  \bibfield  {author} {\bibinfo {author} {\bibfnamefont {M.}~\bibnamefont
  {Fairbairn}}, \bibinfo {author} {\bibfnamefont {T.}~\bibnamefont {Douce}}, \
  and\ \bibinfo {author} {\bibfnamefont {J.}~\bibnamefont {Swift}},\ }\href
  {\doibase 10.1016/j.astropartphys.2013.06.003} {\bibfield  {journal}
  {\bibinfo  {journal} {Astropart. Phys.}\ }\textbf {\bibinfo {volume} {47}},\
  \bibinfo {pages} {45} (\bibinfo {year} {2013})},\ \Eprint
  {http://arxiv.org/abs/1206.2693} {arXiv:1206.2693 [astro-ph.CO]} \BibitemShut
  {NoStop}%
\bibitem [{\citenamefont {Kavanagh}\ and\ \citenamefont
  {Green}(2012)}]{Kavanagh:2012nr}%
  \BibitemOpen
  \bibfield  {author} {\bibinfo {author} {\bibfnamefont {B.~J.}\ \bibnamefont
  {Kavanagh}}\ and\ \bibinfo {author} {\bibfnamefont {A.~M.}\ \bibnamefont
  {Green}},\ }\href {\doibase 10.1103/PhysRevD.86.065027} {\bibfield  {journal}
  {\bibinfo  {journal} {Phys. Rev.}\ }\textbf {\bibinfo {volume} {D86}},\
  \bibinfo {pages} {065027} (\bibinfo {year} {2012})},\ \Eprint
  {http://arxiv.org/abs/1207.2039} {arXiv:1207.2039 [astro-ph.CO]} \BibitemShut
  {NoStop}%
\bibitem [{\citenamefont {Kavanagh}(2014)}]{Kavanagh:2013eya}%
  \BibitemOpen
  \bibfield  {author} {\bibinfo {author} {\bibfnamefont {B.~J.}\ \bibnamefont
  {Kavanagh}},\ }\href {\doibase 10.1103/PhysRevD.89.085026} {\bibfield
  {journal} {\bibinfo  {journal} {Phys. Rev.}\ }\textbf {\bibinfo {volume}
  {D89}},\ \bibinfo {pages} {085026} (\bibinfo {year} {2014})},\ \Eprint
  {http://arxiv.org/abs/1312.1852} {arXiv:1312.1852 [astro-ph.CO]} \BibitemShut
  {NoStop}%
\bibitem [{\citenamefont {Kavanagh}\ and\ \citenamefont
  {Green}(2013)}]{Kavanagh:2013wba}%
  \BibitemOpen
  \bibfield  {author} {\bibinfo {author} {\bibfnamefont {B.~J.}\ \bibnamefont
  {Kavanagh}}\ and\ \bibinfo {author} {\bibfnamefont {A.~M.}\ \bibnamefont
  {Green}},\ }\href {\doibase 10.1103/PhysRevLett.111.031302} {\bibfield
  {journal} {\bibinfo  {journal} {Phys. Rev. Lett.}\ }\textbf {\bibinfo
  {volume} {111}},\ \bibinfo {pages} {031302} (\bibinfo {year} {2013})},\
  \Eprint {http://arxiv.org/abs/1303.6868} {arXiv:1303.6868 [astro-ph.CO]}
  \BibitemShut {NoStop}%
\bibitem [{\citenamefont {Kavanagh}\ \emph {et~al.}(2015)\citenamefont
  {Kavanagh}, \citenamefont {Fornasa},\ and\ \citenamefont
  {Green}}]{Kavanagh:2014rya}%
  \BibitemOpen
  \bibfield  {author} {\bibinfo {author} {\bibfnamefont {B.~J.}\ \bibnamefont
  {Kavanagh}}, \bibinfo {author} {\bibfnamefont {M.}~\bibnamefont {Fornasa}}, \
  and\ \bibinfo {author} {\bibfnamefont {A.~M.}\ \bibnamefont {Green}},\ }\href
  {\doibase 10.1103/PhysRevD.91.103533} {\bibfield  {journal} {\bibinfo
  {journal} {Phys. Rev.}\ }\textbf {\bibinfo {volume} {D91}},\ \bibinfo {pages}
  {103533} (\bibinfo {year} {2015})},\ \Eprint {http://arxiv.org/abs/1410.8051}
  {arXiv:1410.8051 [astro-ph.CO]} \BibitemShut {NoStop}%
\bibitem [{\citenamefont {Benito}\ \emph {et~al.}(2019)\citenamefont {Benito},
  \citenamefont {Cuoco},\ and\ \citenamefont {Iocco}}]{Benito:2019ngh}%
  \BibitemOpen
  \bibfield  {author} {\bibinfo {author} {\bibfnamefont {M.}~\bibnamefont
  {Benito}}, \bibinfo {author} {\bibfnamefont {A.}~\bibnamefont {Cuoco}}, \
  and\ \bibinfo {author} {\bibfnamefont {F.}~\bibnamefont {Iocco}},\ }\href
  {\doibase 10.1088/1475-7516/2019/03/033} {\bibfield  {journal} {\bibinfo
  {journal} {JCAP}\ }\textbf {\bibinfo {volume} {1903}},\ \bibinfo {pages}
  {033} (\bibinfo {year} {2019})},\ \Eprint {http://arxiv.org/abs/1901.02460}
  {arXiv:1901.02460 [astro-ph.GA]} \BibitemShut {NoStop}%
\bibitem [{\citenamefont {{Karukes}}\ \emph {et~al.}(2019)\citenamefont
  {{Karukes}}, \citenamefont {{Benito}}, \citenamefont {{Iocco}}, \citenamefont
  {{Trotta}},\ and\ \citenamefont {{Geringer-Sameth}}}]{2019JCAP...09..046K}%
  \BibitemOpen
  \bibfield  {author} {\bibinfo {author} {\bibfnamefont {E.~V.}\ \bibnamefont
  {{Karukes}}}, \bibinfo {author} {\bibfnamefont {M.}~\bibnamefont {{Benito}}},
  \bibinfo {author} {\bibfnamefont {F.}~\bibnamefont {{Iocco}}}, \bibinfo
  {author} {\bibfnamefont {R.}~\bibnamefont {{Trotta}}}, \ and\ \bibinfo
  {author} {\bibfnamefont {A.}~\bibnamefont {{Geringer-Sameth}}},\ }\href
  {\doibase 10.1088/1475-7516/2019/09/046} {\bibfield  {journal} {\bibinfo
  {journal} {JCAPf}\ }\textbf {\bibinfo {volume} {2019}},\ \bibinfo {eid} {046}
  (\bibinfo {year} {2019})},\ \Eprint {http://arxiv.org/abs/1901.02463}
  {arXiv:1901.02463 [astro-ph.GA]} \BibitemShut {NoStop}%
\bibitem [{\citenamefont {{Pitjev}}\ and\ \citenamefont
  {{Pitjeva}}(2013)}]{2013AstL...39..141P}%
  \BibitemOpen
  \bibfield  {author} {\bibinfo {author} {\bibfnamefont {N.~P.}\ \bibnamefont
  {{Pitjev}}}\ and\ \bibinfo {author} {\bibfnamefont {E.~V.}\ \bibnamefont
  {{Pitjeva}}},\ }\href {\doibase 10.1134/S1063773713020060} {\bibfield
  {journal} {\bibinfo  {journal} {Astronomy Letters}\ }\textbf {\bibinfo
  {volume} {39}},\ \bibinfo {pages} {141} (\bibinfo {year} {2013})},\ \Eprint
  {http://arxiv.org/abs/1306.5534} {arXiv:1306.5534 [astro-ph.EP]} \BibitemShut
  {NoStop}%
\bibitem [{\citenamefont {Read}(2014)}]{Read:2014qva}%
  \BibitemOpen
  \bibfield  {author} {\bibinfo {author} {\bibfnamefont {J.~I.}\ \bibnamefont
  {Read}},\ }\href {\doibase 10.1088/0954-3899/41/6/063101} {\bibfield
  {journal} {\bibinfo  {journal} {J. Phys.}\ }\textbf {\bibinfo {volume}
  {G41}},\ \bibinfo {pages} {063101} (\bibinfo {year} {2014})},\ \Eprint
  {http://arxiv.org/abs/1404.1938} {arXiv:1404.1938 [astro-ph.GA]} \BibitemShut
  {NoStop}%
\bibitem [{\citenamefont {Peter}\ \emph {et~al.}(2013)\citenamefont {Peter},
  \citenamefont {Rocha}, \citenamefont {Bullock},\ and\ \citenamefont
  {Kaplinghat}}]{Peter:2012jh}%
  \BibitemOpen
  \bibfield  {author} {\bibinfo {author} {\bibfnamefont {A.~H.}\ \bibnamefont
  {Peter}}, \bibinfo {author} {\bibfnamefont {M.}~\bibnamefont {Rocha}},
  \bibinfo {author} {\bibfnamefont {J.~S.}\ \bibnamefont {Bullock}}, \ and\
  \bibinfo {author} {\bibfnamefont {M.}~\bibnamefont {Kaplinghat}},\ }\href
  {\doibase 10.1093/mnras/sts535} {\bibfield  {journal} {\bibinfo  {journal}
  {Mon. Not. Roy. Astron. Soc.}\ }\textbf {\bibinfo {volume} {430}},\ \bibinfo
  {pages} {105} (\bibinfo {year} {2013})},\ \Eprint
  {http://arxiv.org/abs/1208.3026} {arXiv:1208.3026 [astro-ph.CO]} \BibitemShut
  {NoStop}%
\bibitem [{\citenamefont {Fan}\ \emph {et~al.}(2013{\natexlab{a}})\citenamefont
  {Fan}, \citenamefont {Katz}, \citenamefont {Randall},\ and\ \citenamefont
  {Reece}}]{Fan:2013tia}%
  \BibitemOpen
  \bibfield  {author} {\bibinfo {author} {\bibfnamefont {J.}~\bibnamefont
  {Fan}}, \bibinfo {author} {\bibfnamefont {A.}~\bibnamefont {Katz}}, \bibinfo
  {author} {\bibfnamefont {L.}~\bibnamefont {Randall}}, \ and\ \bibinfo
  {author} {\bibfnamefont {M.}~\bibnamefont {Reece}},\ }\href {\doibase
  10.1103/PhysRevLett.110.211302} {\bibfield  {journal} {\bibinfo  {journal}
  {Phys. Rev. Lett.}\ }\textbf {\bibinfo {volume} {110}},\ \bibinfo {pages}
  {211302} (\bibinfo {year} {2013}{\natexlab{a}})},\ \Eprint
  {http://arxiv.org/abs/1303.3271} {arXiv:1303.3271 [hep-ph]} \BibitemShut
  {NoStop}%
\bibitem [{\citenamefont {Fan}\ \emph {et~al.}(2013{\natexlab{b}})\citenamefont
  {Fan}, \citenamefont {Katz}, \citenamefont {Randall},\ and\ \citenamefont
  {Reece}}]{Fan:2013yva}%
  \BibitemOpen
  \bibfield  {author} {\bibinfo {author} {\bibfnamefont {J.}~\bibnamefont
  {Fan}}, \bibinfo {author} {\bibfnamefont {A.}~\bibnamefont {Katz}}, \bibinfo
  {author} {\bibfnamefont {L.}~\bibnamefont {Randall}}, \ and\ \bibinfo
  {author} {\bibfnamefont {M.}~\bibnamefont {Reece}},\ }\href {\doibase
  10.1016/j.dark.2013.07.001} {\bibfield  {journal} {\bibinfo  {journal} {Phys.
  Dark Univ.}\ }\textbf {\bibinfo {volume} {2}},\ \bibinfo {pages} {139}
  (\bibinfo {year} {2013}{\natexlab{b}})},\ \Eprint
  {http://arxiv.org/abs/1303.1521} {arXiv:1303.1521 [astro-ph.CO]} \BibitemShut
  {NoStop}%
\bibitem [{\citenamefont {Read}\ \emph {et~al.}(2008)\citenamefont {Read},
  \citenamefont {Lake}, \citenamefont {Agertz},\ and\ \citenamefont
  {Debattista}}]{Read:2008fh}%
  \BibitemOpen
  \bibfield  {author} {\bibinfo {author} {\bibfnamefont {J.}~\bibnamefont
  {Read}}, \bibinfo {author} {\bibfnamefont {G.}~\bibnamefont {Lake}}, \bibinfo
  {author} {\bibfnamefont {O.}~\bibnamefont {Agertz}}, \ and\ \bibinfo {author}
  {\bibfnamefont {V.~P.}\ \bibnamefont {Debattista}},\ }\href {\doibase
  10.1111/j.1365-2966.2008.13643.x} {\bibfield  {journal} {\bibinfo  {journal}
  {Mon. Not. Roy. Astron. Soc.}\ }\textbf {\bibinfo {volume} {389}},\ \bibinfo
  {pages} {1041} (\bibinfo {year} {2008})},\ \Eprint
  {http://arxiv.org/abs/0803.2714} {arXiv:0803.2714 [astro-ph]} \BibitemShut
  {NoStop}%
\bibitem [{\citenamefont {Read}\ \emph {et~al.}(2009)\citenamefont {Read},
  \citenamefont {Mayer}, \citenamefont {Brooks}, \citenamefont {Governato},\
  and\ \citenamefont {Lake}}]{10.1111/j.1365-2966.2009.14757.x}%
  \BibitemOpen
  \bibfield  {author} {\bibinfo {author} {\bibfnamefont {J.~I.}\ \bibnamefont
  {Read}}, \bibinfo {author} {\bibfnamefont {L.}~\bibnamefont {Mayer}},
  \bibinfo {author} {\bibfnamefont {A.~M.}\ \bibnamefont {Brooks}}, \bibinfo
  {author} {\bibfnamefont {F.}~\bibnamefont {Governato}}, \ and\ \bibinfo
  {author} {\bibfnamefont {G.}~\bibnamefont {Lake}},\ }\href {\doibase
  10.1111/j.1365-2966.2009.14757.x} {\bibfield  {journal} {\bibinfo  {journal}
  {Monthly Notices of the Royal Astronomical Society}\ }\textbf {\bibinfo
  {volume} {397}},\ \bibinfo {pages} {44} (\bibinfo {year} {2009})},\ \Eprint
  {http://arxiv.org/abs/https://academic.oup.com/mnras/article-pdf/397/1/44/18429301/mnras0397-0044.pdf}
  {https://academic.oup.com/mnras/article-pdf/397/1/44/18429301/mnras0397-0044.pdf}
  \BibitemShut {NoStop}%
\bibitem [{\citenamefont {Dubinski}(1994)}]{Dubinski:1993df}%
  \BibitemOpen
  \bibfield  {author} {\bibinfo {author} {\bibfnamefont {J.}~\bibnamefont
  {Dubinski}},\ }\href {\doibase 10.1086/174512} {\bibfield  {journal}
  {\bibinfo  {journal} {Astrophys. J.}\ }\textbf {\bibinfo {volume} {431}},\
  \bibinfo {pages} {617} (\bibinfo {year} {1994})},\ \Eprint
  {http://arxiv.org/abs/astro-ph/9309001} {arXiv:astro-ph/9309001} \BibitemShut
  {NoStop}%
\bibitem [{\citenamefont {Kazantzidis}\ \emph {et~al.}(2004)\citenamefont
  {Kazantzidis}, \citenamefont {Kravtsov}, \citenamefont {Zentner},
  \citenamefont {Allgood}, \citenamefont {Nagai},\ and\ \citenamefont
  {Moore}}]{Kazantzidis:2004vu}%
  \BibitemOpen
  \bibfield  {author} {\bibinfo {author} {\bibfnamefont {S.}~\bibnamefont
  {Kazantzidis}}, \bibinfo {author} {\bibfnamefont {A.~V.}\ \bibnamefont
  {Kravtsov}}, \bibinfo {author} {\bibfnamefont {A.~R.}\ \bibnamefont
  {Zentner}}, \bibinfo {author} {\bibfnamefont {B.}~\bibnamefont {Allgood}},
  \bibinfo {author} {\bibfnamefont {D.}~\bibnamefont {Nagai}}, \ and\ \bibinfo
  {author} {\bibfnamefont {B.}~\bibnamefont {Moore}},\ }\href {\doibase
  10.1086/423992} {\bibfield  {journal} {\bibinfo  {journal} {Astrophys. J.
  Lett.}\ }\textbf {\bibinfo {volume} {611}},\ \bibinfo {pages} {L73} (\bibinfo
  {year} {2004})},\ \Eprint {http://arxiv.org/abs/astro-ph/0405189}
  {arXiv:astro-ph/0405189} \BibitemShut {NoStop}%
\bibitem [{\citenamefont {Maccio'}\ \emph {et~al.}(2007)\citenamefont
  {Maccio'}, \citenamefont {Dutton}, \citenamefont {van~den Bosch},
  \citenamefont {Moore}, \citenamefont {Potter},\ and\ \citenamefont
  {Stadel}}]{Maccio:2006wpz}%
  \BibitemOpen
  \bibfield  {author} {\bibinfo {author} {\bibfnamefont {A.~V.}\ \bibnamefont
  {Maccio'}}, \bibinfo {author} {\bibfnamefont {A.~A.}\ \bibnamefont {Dutton}},
  \bibinfo {author} {\bibfnamefont {F.~C.}\ \bibnamefont {van~den Bosch}},
  \bibinfo {author} {\bibfnamefont {B.}~\bibnamefont {Moore}}, \bibinfo
  {author} {\bibfnamefont {D.}~\bibnamefont {Potter}}, \ and\ \bibinfo {author}
  {\bibfnamefont {J.}~\bibnamefont {Stadel}},\ }\href {\doibase
  10.1111/j.1365-2966.2007.11720.x} {\bibfield  {journal} {\bibinfo  {journal}
  {Mon. Not. Roy. Astron. Soc.}\ }\textbf {\bibinfo {volume} {378}},\ \bibinfo
  {pages} {55} (\bibinfo {year} {2007})},\ \Eprint
  {http://arxiv.org/abs/astro-ph/0608157} {arXiv:astro-ph/0608157} \BibitemShut
  {NoStop}%
\bibitem [{\citenamefont {Debattista}\ \emph {et~al.}(2008)\citenamefont
  {Debattista}, \citenamefont {Moore}, \citenamefont {Quinn}, \citenamefont
  {Kazantzidis}, \citenamefont {Maas}, \citenamefont {Mayer}, \citenamefont
  {Read},\ and\ \citenamefont {Stadel}}]{Debattista_2008}%
  \BibitemOpen
  \bibfield  {author} {\bibinfo {author} {\bibfnamefont {V.~P.}\ \bibnamefont
  {Debattista}}, \bibinfo {author} {\bibfnamefont {B.}~\bibnamefont {Moore}},
  \bibinfo {author} {\bibfnamefont {T.}~\bibnamefont {Quinn}}, \bibinfo
  {author} {\bibfnamefont {S.}~\bibnamefont {Kazantzidis}}, \bibinfo {author}
  {\bibfnamefont {R.}~\bibnamefont {Maas}}, \bibinfo {author} {\bibfnamefont
  {L.}~\bibnamefont {Mayer}}, \bibinfo {author} {\bibfnamefont
  {J.}~\bibnamefont {Read}}, \ and\ \bibinfo {author} {\bibfnamefont
  {J.}~\bibnamefont {Stadel}},\ }\href {\doibase 10.1086/587977} {\bibfield
  {journal} {\bibinfo  {journal} {The Astrophysical Journal}\ }\textbf
  {\bibinfo {volume} {681}},\ \bibinfo {pages} {1076} (\bibinfo {year}
  {2008})}\BibitemShut {NoStop}%
\bibitem [{\citenamefont {Hellwing}\ \emph {et~al.}(2013)\citenamefont
  {Hellwing}, \citenamefont {Cautun}, \citenamefont {Knebe}, \citenamefont
  {Juszkiewicz},\ and\ \citenamefont {Knollmann}}]{Hellwing:2011ne}%
  \BibitemOpen
  \bibfield  {author} {\bibinfo {author} {\bibfnamefont {W.~A.}\ \bibnamefont
  {Hellwing}}, \bibinfo {author} {\bibfnamefont {M.}~\bibnamefont {Cautun}},
  \bibinfo {author} {\bibfnamefont {A.}~\bibnamefont {Knebe}}, \bibinfo
  {author} {\bibfnamefont {R.}~\bibnamefont {Juszkiewicz}}, \ and\ \bibinfo
  {author} {\bibfnamefont {S.}~\bibnamefont {Knollmann}},\ }\href {\doibase
  10.1088/1475-7516/2013/10/012} {\bibfield  {journal} {\bibinfo  {journal}
  {JCAP}\ }\textbf {\bibinfo {volume} {10}},\ \bibinfo {pages} {012} (\bibinfo
  {year} {2013})},\ \Eprint {http://arxiv.org/abs/1111.7257} {arXiv:1111.7257
  [astro-ph.CO]} \BibitemShut {NoStop}%
\bibitem [{\citenamefont {{Chua}}\ \emph {et~al.}(2019)\citenamefont {{Chua}},
  \citenamefont {{Pillepich}}, \citenamefont {{Vogelsberger}},\ and\
  \citenamefont {{Hernquist}}}]{2019MNRAS.484..476C}%
  \BibitemOpen
  \bibfield  {author} {\bibinfo {author} {\bibfnamefont {K.~T.~E.}\
  \bibnamefont {{Chua}}}, \bibinfo {author} {\bibfnamefont {A.}~\bibnamefont
  {{Pillepich}}}, \bibinfo {author} {\bibfnamefont {M.}~\bibnamefont
  {{Vogelsberger}}}, \ and\ \bibinfo {author} {\bibfnamefont {L.}~\bibnamefont
  {{Hernquist}}},\ }\href {\doibase 10.1093/mnras/sty3531} {\bibfield
  {journal} {\bibinfo  {journal} {Monthly Notices of the Royal Astronomical
  Society}\ }\textbf {\bibinfo {volume} {484}},\ \bibinfo {pages} {476}
  (\bibinfo {year} {2019})},\ \Eprint {http://arxiv.org/abs/1809.07255}
  {arXiv:1809.07255 [astro-ph.GA]} \BibitemShut {NoStop}%
\bibitem [{\citenamefont {Ibata}\ \emph {et~al.}(2001)\citenamefont {Ibata},
  \citenamefont {Lewis}, \citenamefont {Irwin}, \citenamefont {Totten},\ and\
  \citenamefont {Quinn}}]{Ibata:2000pu}%
  \BibitemOpen
  \bibfield  {author} {\bibinfo {author} {\bibfnamefont {R.}~\bibnamefont
  {Ibata}}, \bibinfo {author} {\bibfnamefont {G.~F.}\ \bibnamefont {Lewis}},
  \bibinfo {author} {\bibfnamefont {M.}~\bibnamefont {Irwin}}, \bibinfo
  {author} {\bibfnamefont {E.}~\bibnamefont {Totten}}, \ and\ \bibinfo {author}
  {\bibfnamefont {T.~R.}\ \bibnamefont {Quinn}},\ }\href {\doibase
  10.1086/320060} {\bibfield  {journal} {\bibinfo  {journal} {Astrophys. J.}\
  }\textbf {\bibinfo {volume} {551}},\ \bibinfo {pages} {294} (\bibinfo {year}
  {2001})},\ \Eprint {http://arxiv.org/abs/astro-ph/0004011}
  {arXiv:astro-ph/0004011} \BibitemShut {NoStop}%
\bibitem [{\citenamefont {{Law}}\ and\ \citenamefont
  {{Majewski}}(2010)}]{2010ApJ...714..229L}%
  \BibitemOpen
  \bibfield  {author} {\bibinfo {author} {\bibfnamefont {D.~R.}\ \bibnamefont
  {{Law}}}\ and\ \bibinfo {author} {\bibfnamefont {S.~R.}\ \bibnamefont
  {{Majewski}}},\ }\href {\doibase 10.1088/0004-637X/714/1/229} {\bibfield
  {journal} {\bibinfo  {journal} {\apj}\ }\textbf {\bibinfo {volume} {714}},\
  \bibinfo {pages} {229} (\bibinfo {year} {2010})},\ \Eprint
  {http://arxiv.org/abs/1003.1132} {arXiv:1003.1132 [astro-ph.GA]} \BibitemShut
  {NoStop}%
\bibitem [{\citenamefont {Bovy}\ \emph {et~al.}(2016)\citenamefont {Bovy},
  \citenamefont {Bahmanyar}, \citenamefont {Fritz},\ and\ \citenamefont
  {Kallivayalil}}]{Bovy:2016chl}%
  \BibitemOpen
  \bibfield  {author} {\bibinfo {author} {\bibfnamefont {J.}~\bibnamefont
  {Bovy}}, \bibinfo {author} {\bibfnamefont {A.}~\bibnamefont {Bahmanyar}},
  \bibinfo {author} {\bibfnamefont {T.~K.}\ \bibnamefont {Fritz}}, \ and\
  \bibinfo {author} {\bibfnamefont {N.}~\bibnamefont {Kallivayalil}},\ }\href
  {\doibase 10.3847/1538-4357/833/1/31} {\bibfield  {journal} {\bibinfo
  {journal} {Astrophys. J.}\ }\textbf {\bibinfo {volume} {833}},\ \bibinfo
  {pages} {31} (\bibinfo {year} {2016})},\ \Eprint
  {http://arxiv.org/abs/1609.01298} {arXiv:1609.01298 [astro-ph.GA]}
  \BibitemShut {NoStop}%
\bibitem [{\citenamefont {Wegg}\ \emph {et~al.}(2019)\citenamefont {Wegg},
  \citenamefont {Gerhard},\ and\ \citenamefont {Bieth}}]{Wegg2019May}%
  \BibitemOpen
  \bibfield  {author} {\bibinfo {author} {\bibfnamefont {C.}~\bibnamefont
  {Wegg}}, \bibinfo {author} {\bibfnamefont {O.}~\bibnamefont {Gerhard}}, \
  and\ \bibinfo {author} {\bibfnamefont {M.}~\bibnamefont {Bieth}},\ }\href
  {\doibase 10.1093/mnras/stz572} {\bibfield  {journal} {\bibinfo  {journal}
  {Mon. Not. R. Astron. Soc.}\ }\textbf {\bibinfo {volume} {485}},\ \bibinfo
  {pages} {3296} (\bibinfo {year} {2019})},\ \Eprint
  {http://arxiv.org/abs/arXiv:1806.09635} {arXiv:1806.09635} \BibitemShut
  {NoStop}%
\bibitem [{\citenamefont {Dai}\ \emph {et~al.}(2018)\citenamefont {Dai},
  \citenamefont {Robertson},\ and\ \citenamefont {Madau}}]{Dai:2018ypv}%
  \BibitemOpen
  \bibfield  {author} {\bibinfo {author} {\bibfnamefont {B.}~\bibnamefont
  {Dai}}, \bibinfo {author} {\bibfnamefont {B.~E.}\ \bibnamefont {Robertson}},
  \ and\ \bibinfo {author} {\bibfnamefont {P.}~\bibnamefont {Madau}},\ }\href
  {\doibase 10.3847/1538-4357/aabb06} {\bibfield  {journal} {\bibinfo
  {journal} {Astrophys. J.}\ }\textbf {\bibinfo {volume} {858}},\ \bibinfo
  {pages} {73} (\bibinfo {year} {2018})},\ \Eprint
  {http://arxiv.org/abs/1804.00669} {arXiv:1804.00669 [astro-ph.GA]}
  \BibitemShut {NoStop}%
\bibitem [{\citenamefont {Prada}\ \emph {et~al.}(2019)\citenamefont {Prada},
  \citenamefont {Forero-Romero}, \citenamefont {Grand}, \citenamefont
  {Pakmor},\ and\ \citenamefont {Springel}}]{Prada2019Dec}%
  \BibitemOpen
  \bibfield  {author} {\bibinfo {author} {\bibfnamefont {J.}~\bibnamefont
  {Prada}}, \bibinfo {author} {\bibfnamefont {J.~E.}\ \bibnamefont
  {Forero-Romero}}, \bibinfo {author} {\bibfnamefont {R.~J.~J.}\ \bibnamefont
  {Grand}}, \bibinfo {author} {\bibfnamefont {R.}~\bibnamefont {Pakmor}}, \
  and\ \bibinfo {author} {\bibfnamefont {V.}~\bibnamefont {Springel}},\ }\href
  {\doibase 10.1093/mnras/stz2873} {\bibfield  {journal} {\bibinfo  {journal}
  {Mon. Not. R. Astron. Soc.}\ }\textbf {\bibinfo {volume} {490}},\ \bibinfo
  {pages} {4877} (\bibinfo {year} {2019})},\ \Eprint
  {http://arxiv.org/abs/arXiv:1910.04045} {arXiv:1910.04045} \BibitemShut
  {NoStop}%
\bibitem [{\citenamefont {Poole-McKenzie}\ \emph {et~al.}(2020)\citenamefont
  {Poole-McKenzie}, \citenamefont {Font}, \citenamefont {Boxer}, \citenamefont
  {McCarthy}, \citenamefont {Burdin}, \citenamefont {Stafford},\ and\
  \citenamefont {Brown}}]{Poole-McKenzie:2020dbo}%
  \BibitemOpen
  \bibfield  {author} {\bibinfo {author} {\bibfnamefont {R.}~\bibnamefont
  {Poole-McKenzie}}, \bibinfo {author} {\bibfnamefont {A.~S.}\ \bibnamefont
  {Font}}, \bibinfo {author} {\bibfnamefont {B.}~\bibnamefont {Boxer}},
  \bibinfo {author} {\bibfnamefont {I.~G.}\ \bibnamefont {McCarthy}}, \bibinfo
  {author} {\bibfnamefont {S.}~\bibnamefont {Burdin}}, \bibinfo {author}
  {\bibfnamefont {S.~G.}\ \bibnamefont {Stafford}}, \ and\ \bibinfo {author}
  {\bibfnamefont {S.~T.}\ \bibnamefont {Brown}},\ }\href {\doibase
  10.1088/1475-7516/2020/11/016} {\bibfield  {journal} {\bibinfo  {journal}
  {JCAP}\ }\textbf {\bibinfo {volume} {11}},\ \bibinfo {pages} {016} (\bibinfo
  {year} {2020})},\ \Eprint {http://arxiv.org/abs/2006.15159} {arXiv:2006.15159
  [astro-ph.CO]} \BibitemShut {NoStop}%
\bibitem [{\citenamefont {Cataldi}\ \emph {et~al.}(2020)\citenamefont
  {Cataldi}, \citenamefont {Pedrosa}, \citenamefont {Tissera},\ and\
  \citenamefont {Artale}}]{Cataldi:2020kvs}%
  \BibitemOpen
  \bibfield  {author} {\bibinfo {author} {\bibfnamefont {P.}~\bibnamefont
  {Cataldi}}, \bibinfo {author} {\bibfnamefont {S.}~\bibnamefont {Pedrosa}},
  \bibinfo {author} {\bibfnamefont {P.}~\bibnamefont {Tissera}}, \ and\
  \bibinfo {author} {\bibfnamefont {C.}~\bibnamefont {Artale}},\ }\href@noop {}
  {\  (\bibinfo {year} {2020})},\ \Eprint {http://arxiv.org/abs/2008.02404}
  {arXiv:2008.02404 [astro-ph.GA]} \BibitemShut {NoStop}%
\bibitem [{\citenamefont {Emami}\ \emph {et~al.}(2020)\citenamefont {Emami}
  \emph {et~al.}}]{Emami:2020cwt}%
  \BibitemOpen
  \bibfield  {author} {\bibinfo {author} {\bibfnamefont {R.}~\bibnamefont
  {Emami}} \emph {et~al.},\ }\href@noop {} {\  (\bibinfo {year} {2020})},\
  \Eprint {http://arxiv.org/abs/2009.09220} {arXiv:2009.09220 [astro-ph.GA]}
  \BibitemShut {NoStop}%
\bibitem [{\citenamefont {Drukier}\ and\ \citenamefont
  {Stodolsky}(1984)}]{Drukier:1983gj}%
  \BibitemOpen
  \bibfield  {author} {\bibinfo {author} {\bibfnamefont {A.}~\bibnamefont
  {Drukier}}\ and\ \bibinfo {author} {\bibfnamefont {L.}~\bibnamefont
  {Stodolsky}},\ }\href {\doibase 10.1103/PhysRevD.30.2295} {\bibfield
  {journal} {\bibinfo  {journal} {Phys. Rev. D}\ }\textbf {\bibinfo {volume}
  {30}},\ \bibinfo {pages} {2295} (\bibinfo {year} {1984})}\BibitemShut
  {NoStop}%
\bibitem [{\citenamefont {Goodman}\ and\ \citenamefont
  {Witten}(1985)}]{Goodman:1984dc}%
  \BibitemOpen
  \bibfield  {author} {\bibinfo {author} {\bibfnamefont {M.~W.}\ \bibnamefont
  {Goodman}}\ and\ \bibinfo {author} {\bibfnamefont {E.}~\bibnamefont
  {Witten}},\ }\href {\doibase 10.1103/PhysRevD.31.3059} {\bibfield  {journal}
  {\bibinfo  {journal} {Phys. Rev. D}\ }\textbf {\bibinfo {volume} {31}},\
  \bibinfo {pages} {3059} (\bibinfo {year} {1985})}\BibitemShut {NoStop}%
\bibitem [{\citenamefont {Drukier}\ \emph {et~al.}(1986)\citenamefont
  {Drukier}, \citenamefont {Freese},\ and\ \citenamefont
  {Spergel}}]{Drukier:1986tm}%
  \BibitemOpen
  \bibfield  {author} {\bibinfo {author} {\bibfnamefont {A.}~\bibnamefont
  {Drukier}}, \bibinfo {author} {\bibfnamefont {K.}~\bibnamefont {Freese}}, \
  and\ \bibinfo {author} {\bibfnamefont {D.}~\bibnamefont {Spergel}},\ }\href
  {\doibase 10.1103/PhysRevD.33.3495} {\bibfield  {journal} {\bibinfo
  {journal} {Phys. Rev. D}\ }\textbf {\bibinfo {volume} {33}},\ \bibinfo
  {pages} {3495} (\bibinfo {year} {1986})}\BibitemShut {NoStop}%
\bibitem [{\citenamefont {Gould}\ \emph {et~al.}(1989)\citenamefont {Gould},
  \citenamefont {Frieman},\ and\ \citenamefont {Freese}}]{Gould:1988eq}%
  \BibitemOpen
  \bibfield  {author} {\bibinfo {author} {\bibfnamefont {A.}~\bibnamefont
  {Gould}}, \bibinfo {author} {\bibfnamefont {J.~A.}\ \bibnamefont {Frieman}},
  \ and\ \bibinfo {author} {\bibfnamefont {K.}~\bibnamefont {Freese}},\ }\href
  {\doibase 10.1103/PhysRevD.39.1029} {\bibfield  {journal} {\bibinfo
  {journal} {Phys. Rev. D}\ }\textbf {\bibinfo {volume} {39}},\ \bibinfo
  {pages} {1029} (\bibinfo {year} {1989})}\BibitemShut {NoStop}%
\bibitem [{\citenamefont {Collar}\ and\ \citenamefont
  {Avignone}(1993)}]{Collar:1993ss}%
  \BibitemOpen
  \bibfield  {author} {\bibinfo {author} {\bibfnamefont {J.~I.}\ \bibnamefont
  {Collar}}\ and\ \bibinfo {author} {\bibfnamefont {F.~T.}\ \bibnamefont
  {Avignone}, \bibfnamefont {III}},\ }\href {\doibase 10.1103/PhysRevD.47.5238}
  {\bibfield  {journal} {\bibinfo  {journal} {Phys. Rev.}\ }\textbf {\bibinfo
  {volume} {D47}},\ \bibinfo {pages} {5238} (\bibinfo {year}
  {1993})}\BibitemShut {NoStop}%
\bibitem [{\citenamefont {Hasenbalg}\ \emph {et~al.}(1997)\citenamefont
  {Hasenbalg}, \citenamefont {Abriola}, \citenamefont {Avignone}, \citenamefont
  {Collar}, \citenamefont {Di~Gregorio}, \citenamefont {Gattone}, \citenamefont
  {Huck}, \citenamefont {Tomasi},\ and\ \citenamefont
  {Urteaga}}]{Hasenbalg:1997hs}%
  \BibitemOpen
  \bibfield  {author} {\bibinfo {author} {\bibfnamefont {F.}~\bibnamefont
  {Hasenbalg}}, \bibinfo {author} {\bibfnamefont {D.}~\bibnamefont {Abriola}},
  \bibinfo {author} {\bibfnamefont {F.~T.}\ \bibnamefont {Avignone}}, \bibinfo
  {author} {\bibfnamefont {J.~I.}\ \bibnamefont {Collar}}, \bibinfo {author}
  {\bibfnamefont {D.~E.}\ \bibnamefont {Di~Gregorio}}, \bibinfo {author}
  {\bibfnamefont {A.~O.}\ \bibnamefont {Gattone}}, \bibinfo {author}
  {\bibfnamefont {H.}~\bibnamefont {Huck}}, \bibinfo {author} {\bibfnamefont
  {D.}~\bibnamefont {Tomasi}}, \ and\ \bibinfo {author} {\bibfnamefont
  {I.}~\bibnamefont {Urteaga}},\ }\href {\doibase 10.1103/PhysRevD.55.7350}
  {\bibfield  {journal} {\bibinfo  {journal} {Phys. Rev.}\ }\textbf {\bibinfo
  {volume} {D55}},\ \bibinfo {pages} {7350} (\bibinfo {year} {1997})},\ \Eprint
  {http://arxiv.org/abs/astro-ph/9702165} {arXiv:astro-ph/9702165 [astro-ph]}
  \BibitemShut {NoStop}%
\bibitem [{\citenamefont {Kavanagh}\ \emph {et~al.}(2017)\citenamefont
  {Kavanagh}, \citenamefont {Catena},\ and\ \citenamefont
  {Kouvaris}}]{Kavanagh:2016pyr}%
  \BibitemOpen
  \bibfield  {author} {\bibinfo {author} {\bibfnamefont {B.~J.}\ \bibnamefont
  {Kavanagh}}, \bibinfo {author} {\bibfnamefont {R.}~\bibnamefont {Catena}}, \
  and\ \bibinfo {author} {\bibfnamefont {C.}~\bibnamefont {Kouvaris}},\ }\href
  {\doibase 10.1088/1475-7516/2017/01/012} {\bibfield  {journal} {\bibinfo
  {journal} {JCAP}\ }\textbf {\bibinfo {volume} {1701}},\ \bibinfo {pages}
  {012} (\bibinfo {year} {2017})},\ \Eprint {http://arxiv.org/abs/1611.05453}
  {arXiv:1611.05453 [hep-ph]} \BibitemShut {NoStop}%
\bibitem [{\citenamefont {Emken}\ and\ \citenamefont
  {Kouvaris}(2017)}]{Emken:2017qmp}%
  \BibitemOpen
  \bibfield  {author} {\bibinfo {author} {\bibfnamefont {T.}~\bibnamefont
  {Emken}}\ and\ \bibinfo {author} {\bibfnamefont {C.}~\bibnamefont
  {Kouvaris}},\ }\href {\doibase 10.1088/1475-7516/2017/10/031} {\bibfield
  {journal} {\bibinfo  {journal} {JCAP}\ }\textbf {\bibinfo {volume} {1710}},\
  \bibinfo {pages} {031} (\bibinfo {year} {2017})},\ \Eprint
  {http://arxiv.org/abs/1706.02249} {arXiv:1706.02249 [hep-ph]} \BibitemShut
  {NoStop}%
\bibitem [{\citenamefont {Aartsen}\ \emph {et~al.}(2017)\citenamefont {Aartsen}
  \emph {et~al.}}]{Aartsen:2017kpd}%
  \BibitemOpen
  \bibfield  {author} {\bibinfo {author} {\bibfnamefont {M.}~\bibnamefont
  {Aartsen}} \emph {et~al.} (\bibinfo {collaboration} {IceCube}),\ }\href
  {\doibase 10.1038/nature24459} {\bibfield  {journal} {\bibinfo  {journal}
  {Nature}\ }\textbf {\bibinfo {volume} {551}},\ \bibinfo {pages} {596}
  (\bibinfo {year} {2017})},\ \Eprint {http://arxiv.org/abs/1711.08119}
  {arXiv:1711.08119 [hep-ex]} \BibitemShut {NoStop}%
\bibitem [{\citenamefont {Bramante}\ \emph {et~al.}(2018)\citenamefont
  {Bramante}, \citenamefont {Broerman}, \citenamefont {Lang},\ and\
  \citenamefont {Raj}}]{Bramante:2018qbc}%
  \BibitemOpen
  \bibfield  {author} {\bibinfo {author} {\bibfnamefont {J.}~\bibnamefont
  {Bramante}}, \bibinfo {author} {\bibfnamefont {B.}~\bibnamefont {Broerman}},
  \bibinfo {author} {\bibfnamefont {R.~F.}\ \bibnamefont {Lang}}, \ and\
  \bibinfo {author} {\bibfnamefont {N.}~\bibnamefont {Raj}},\ }\href {\doibase
  10.1103/PhysRevD.98.083516} {\bibfield  {journal} {\bibinfo  {journal} {Phys.
  Rev.}\ }\textbf {\bibinfo {volume} {D98}},\ \bibinfo {pages} {083516}
  (\bibinfo {year} {2018})},\ \Eprint {http://arxiv.org/abs/1803.08044}
  {arXiv:1803.08044 [hep-ph]} \BibitemShut {NoStop}%
\bibitem [{\citenamefont {Essig}\ \emph {et~al.}(2012)\citenamefont {Essig},
  \citenamefont {Mardon},\ and\ \citenamefont {Volansky}}]{Essig:2011nj}%
  \BibitemOpen
  \bibfield  {author} {\bibinfo {author} {\bibfnamefont {R.}~\bibnamefont
  {Essig}}, \bibinfo {author} {\bibfnamefont {J.}~\bibnamefont {Mardon}}, \
  and\ \bibinfo {author} {\bibfnamefont {T.}~\bibnamefont {Volansky}},\ }\href
  {\doibase 10.1103/PhysRevD.85.076007} {\bibfield  {journal} {\bibinfo
  {journal} {Phys. Rev. D}\ }\textbf {\bibinfo {volume} {85}},\ \bibinfo
  {pages} {076007} (\bibinfo {year} {2012})},\ \Eprint
  {http://arxiv.org/abs/1108.5383} {arXiv:1108.5383 [hep-ph]} \BibitemShut
  {NoStop}%
\bibitem [{\citenamefont {Knapen}\ \emph {et~al.}(2017)\citenamefont {Knapen},
  \citenamefont {Lin},\ and\ \citenamefont {Zurek}}]{Knapen:2016cue}%
  \BibitemOpen
  \bibfield  {author} {\bibinfo {author} {\bibfnamefont {S.}~\bibnamefont
  {Knapen}}, \bibinfo {author} {\bibfnamefont {T.}~\bibnamefont {Lin}}, \ and\
  \bibinfo {author} {\bibfnamefont {K.~M.}\ \bibnamefont {Zurek}},\ }\href
  {\doibase 10.1103/PhysRevD.95.056019} {\bibfield  {journal} {\bibinfo
  {journal} {Phys. Rev. D}\ }\textbf {\bibinfo {volume} {95}},\ \bibinfo
  {pages} {056019} (\bibinfo {year} {2017})},\ \Eprint
  {http://arxiv.org/abs/1611.06228} {arXiv:1611.06228 [hep-ph]} \BibitemShut
  {NoStop}%
\bibitem [{\citenamefont {Hochberg}\ \emph {et~al.}(2018)\citenamefont
  {Hochberg}, \citenamefont {Kahn}, \citenamefont {Lisanti}, \citenamefont
  {Zurek}, \citenamefont {Grushin}, \citenamefont {Ilan}, \citenamefont
  {Griffin}, \citenamefont {Liu}, \citenamefont {Weber},\ and\ \citenamefont
  {Neaton}}]{Hochberg:2017wce}%
  \BibitemOpen
  \bibfield  {author} {\bibinfo {author} {\bibfnamefont {Y.}~\bibnamefont
  {Hochberg}}, \bibinfo {author} {\bibfnamefont {Y.}~\bibnamefont {Kahn}},
  \bibinfo {author} {\bibfnamefont {M.}~\bibnamefont {Lisanti}}, \bibinfo
  {author} {\bibfnamefont {K.~M.}\ \bibnamefont {Zurek}}, \bibinfo {author}
  {\bibfnamefont {A.~G.}\ \bibnamefont {Grushin}}, \bibinfo {author}
  {\bibfnamefont {R.}~\bibnamefont {Ilan}}, \bibinfo {author} {\bibfnamefont
  {S.~M.}\ \bibnamefont {Griffin}}, \bibinfo {author} {\bibfnamefont {Z.-F.}\
  \bibnamefont {Liu}}, \bibinfo {author} {\bibfnamefont {S.~F.}\ \bibnamefont
  {Weber}}, \ and\ \bibinfo {author} {\bibfnamefont {J.~B.}\ \bibnamefont
  {Neaton}},\ }\href {\doibase 10.1103/PhysRevD.97.015004} {\bibfield
  {journal} {\bibinfo  {journal} {Phys. Rev. D}\ }\textbf {\bibinfo {volume}
  {97}},\ \bibinfo {pages} {015004} (\bibinfo {year} {2018})},\ \Eprint
  {http://arxiv.org/abs/1708.08929} {arXiv:1708.08929 [hep-ph]} \BibitemShut
  {NoStop}%
\bibitem [{\citenamefont {Trickle}\ \emph {et~al.}(2020)\citenamefont
  {Trickle}, \citenamefont {Zhang},\ and\ \citenamefont
  {Zurek}}]{Trickle:2019ovy}%
  \BibitemOpen
  \bibfield  {author} {\bibinfo {author} {\bibfnamefont {T.}~\bibnamefont
  {Trickle}}, \bibinfo {author} {\bibfnamefont {Z.}~\bibnamefont {Zhang}}, \
  and\ \bibinfo {author} {\bibfnamefont {K.~M.}\ \bibnamefont {Zurek}},\ }\href
  {\doibase 10.1103/PhysRevLett.124.201801} {\bibfield  {journal} {\bibinfo
  {journal} {Phys. Rev. Lett.}\ }\textbf {\bibinfo {volume} {124}},\ \bibinfo
  {pages} {201801} (\bibinfo {year} {2020})},\ \Eprint
  {http://arxiv.org/abs/1905.13744} {arXiv:1905.13744 [hep-ph]} \BibitemShut
  {NoStop}%
\bibitem [{\citenamefont {Hochberg}\ \emph {et~al.}(2019)\citenamefont
  {Hochberg}, \citenamefont {Charaev}, \citenamefont {Nam}, \citenamefont
  {Verma}, \citenamefont {Colangelo},\ and\ \citenamefont
  {Berggren}}]{Hochberg:2019cyy}%
  \BibitemOpen
  \bibfield  {author} {\bibinfo {author} {\bibfnamefont {Y.}~\bibnamefont
  {Hochberg}}, \bibinfo {author} {\bibfnamefont {I.}~\bibnamefont {Charaev}},
  \bibinfo {author} {\bibfnamefont {S.-W.}\ \bibnamefont {Nam}}, \bibinfo
  {author} {\bibfnamefont {V.}~\bibnamefont {Verma}}, \bibinfo {author}
  {\bibfnamefont {M.}~\bibnamefont {Colangelo}}, \ and\ \bibinfo {author}
  {\bibfnamefont {K.~K.}\ \bibnamefont {Berggren}},\ }\href {\doibase
  10.1103/PhysRevLett.123.151802} {\bibfield  {journal} {\bibinfo  {journal}
  {Phys. Rev. Lett.}\ }\textbf {\bibinfo {volume} {123}},\ \bibinfo {pages}
  {151802} (\bibinfo {year} {2019})},\ \Eprint
  {http://arxiv.org/abs/1903.05101} {arXiv:1903.05101 [hep-ph]} \BibitemShut
  {NoStop}%
\bibitem [{\citenamefont {Bertone}\ and\ \citenamefont
  {Tait}(2018)}]{Bertone:2018xtm}%
  \BibitemOpen
  \bibfield  {author} {\bibinfo {author} {\bibfnamefont {G.}~\bibnamefont
  {Bertone}}\ and\ \bibinfo {author} {\bibfnamefont {M.~P.}\ \bibnamefont
  {Tait}, \bibfnamefont {Tim}},\ }\href {\doibase 10.1038/s41586-018-0542-z}
  {\bibfield  {journal} {\bibinfo  {journal} {Nature}\ }\textbf {\bibinfo
  {volume} {562}},\ \bibinfo {pages} {51} (\bibinfo {year} {2018})},\ \Eprint
  {http://arxiv.org/abs/1810.01668} {arXiv:1810.01668 [astro-ph.CO]}
  \BibitemShut {NoStop}%
\bibitem [{\citenamefont {Emken}\ and\ \citenamefont
  {Kouvaris}(2020)}]{Emken2017}%
  \BibitemOpen
  \bibfield  {author} {\bibinfo {author} {\bibfnamefont {T.}~\bibnamefont
  {Emken}}\ and\ \bibinfo {author} {\bibfnamefont {C.}~\bibnamefont
  {Kouvaris}},\ }\href {\doibase 10.5281/zenodo.3726878} {\enquote {\bibinfo
  {title} {{Dark Matter Simulation Code for Underground Scatterings~(DaMaSCUS)
  [Code, v1.1]}},}\ }\bibinfo {howpublished}
  {\url{https://github.com/temken/damascus},
  \href{https://doi.org/10.5281/zenodo.3726878}{DOI:10.5281/zenodo.3726878}}
  (\bibinfo {year} {2017-2020})\BibitemShut {NoStop}%
\bibitem [{\citenamefont {Kavanagh}\ and\ \citenamefont
  {Catena}(2020)}]{LikelihoodCode}%
  \BibitemOpen
  \bibfield  {author} {\bibinfo {author} {\bibfnamefont {B.~J.}\ \bibnamefont
  {Kavanagh}}\ and\ \bibinfo {author} {\bibfnamefont {R.}~\bibnamefont
  {Catena}},\ }\href {\doibase 10.5281/zenodo.3725882} {\enquote {\bibinfo
  {title} {{EarthScatterLikelihood [Code, v1.1]}},}\ }\bibinfo {howpublished}
  {\url{https://github.com/bradkav/EarthScatterLikelihood},
  \href{https://doi.org/10.5281/zenodo.3725882}{DOI:10.5281/zenodo.3725882}}
  (\bibinfo {year} {2020})\BibitemShut {NoStop}%
\bibitem [{\citenamefont {Lewin}\ and\ \citenamefont
  {Smith}(1996)}]{Lewin:1995rx}%
  \BibitemOpen
  \bibfield  {author} {\bibinfo {author} {\bibfnamefont {J.~D.}\ \bibnamefont
  {Lewin}}\ and\ \bibinfo {author} {\bibfnamefont {P.~F.}\ \bibnamefont
  {Smith}},\ }\href {\doibase 10.1016/S0927-6505(96)00047-3} {\bibfield
  {journal} {\bibinfo  {journal} {Astropart. Phys.}\ }\textbf {\bibinfo
  {volume} {6}},\ \bibinfo {pages} {87} (\bibinfo {year} {1996})}\BibitemShut
  {NoStop}%
\bibitem [{\citenamefont {Cerdeno}\ and\ \citenamefont
  {Green}(2010)}]{Cerdeno:2010jj}%
  \BibitemOpen
  \bibfield  {author} {\bibinfo {author} {\bibfnamefont {D.~G.}\ \bibnamefont
  {Cerdeno}}\ and\ \bibinfo {author} {\bibfnamefont {A.~M.}\ \bibnamefont
  {Green}},\ }\enquote {\bibinfo {title} {{Direct detection of WIMPs}},}\ in\
  \href@noop {} {\emph {\bibinfo {booktitle} {Particle Dark Matter:
  Observations, Models and Searches}}},\ \bibinfo {editor} {edited by\ \bibinfo
  {editor} {\bibfnamefont {G.}~\bibnamefont {Bertone}}}\ (\bibinfo {year}
  {2010})\ pp.\ \bibinfo {pages} {347--369},\ \Eprint
  {http://arxiv.org/abs/1002.1912} {arXiv:1002.1912 [astro-ph.CO]} \BibitemShut
  {NoStop}%
\bibitem [{\citenamefont {Green}(2012)}]{Green:2011bv}%
  \BibitemOpen
  \bibfield  {author} {\bibinfo {author} {\bibfnamefont {A.~M.}\ \bibnamefont
  {Green}},\ }\href {\doibase 10.1142/S0217732312300042} {\bibfield  {journal}
  {\bibinfo  {journal} {Mod. Phys. Lett.}\ }\textbf {\bibinfo {volume} {A27}},\
  \bibinfo {pages} {1230004} (\bibinfo {year} {2012})},\ \Eprint
  {http://arxiv.org/abs/1112.0524} {arXiv:1112.0524 [astro-ph.CO]} \BibitemShut
  {NoStop}%
\bibitem [{\citenamefont {Green}(2017)}]{Green:2017odb}%
  \BibitemOpen
  \bibfield  {author} {\bibinfo {author} {\bibfnamefont {A.~M.}\ \bibnamefont
  {Green}},\ }\href {\doibase 10.1088/1361-6471/aa7819} {\bibfield  {journal}
  {\bibinfo  {journal} {J. Phys.}\ }\textbf {\bibinfo {volume} {G44}},\
  \bibinfo {pages} {084001} (\bibinfo {year} {2017})},\ \Eprint
  {http://arxiv.org/abs/1703.10102} {arXiv:1703.10102 [astro-ph.CO]}
  \BibitemShut {NoStop}%
\bibitem [{\citenamefont {Smith}\ \emph {et~al.}(2007)\citenamefont {Smith}
  \emph {et~al.}}]{Smith:2006ym}%
  \BibitemOpen
  \bibfield  {author} {\bibinfo {author} {\bibfnamefont {M.~C.}\ \bibnamefont
  {Smith}} \emph {et~al.},\ }\href {\doibase 10.1111/j.1365-2966.2007.11964.x}
  {\bibfield  {journal} {\bibinfo  {journal} {Mon. Not. Roy. Astron. Soc.}\
  }\textbf {\bibinfo {volume} {379}},\ \bibinfo {pages} {755} (\bibinfo {year}
  {2007})},\ \Eprint {http://arxiv.org/abs/astro-ph/0611671}
  {arXiv:astro-ph/0611671 [astro-ph]} \BibitemShut {NoStop}%
\bibitem [{\citenamefont {Piffl}\ \emph {et~al.}(2014)\citenamefont {Piffl}
  \emph {et~al.}}]{Piffl:2013mla}%
  \BibitemOpen
  \bibfield  {author} {\bibinfo {author} {\bibfnamefont {T.}~\bibnamefont
  {Piffl}} \emph {et~al.},\ }\href {\doibase 10.1051/0004-6361/201322531}
  {\bibfield  {journal} {\bibinfo  {journal} {Astron. Astrophys.}\ }\textbf
  {\bibinfo {volume} {562}},\ \bibinfo {pages} {A91} (\bibinfo {year}
  {2014})},\ \Eprint {http://arxiv.org/abs/1309.4293} {arXiv:1309.4293
  [astro-ph.GA]} \BibitemShut {NoStop}%
\bibitem [{\citenamefont {Angloher}\ \emph
  {et~al.}(2017{\natexlab{a}})\citenamefont {Angloher} \emph
  {et~al.}}]{Angloher:2017zkf}%
  \BibitemOpen
  \bibfield  {author} {\bibinfo {author} {\bibfnamefont {G.}~\bibnamefont
  {Angloher}} \emph {et~al.} (\bibinfo {collaboration} {CRESST}),\ }\href@noop
  {} {\  (\bibinfo {year} {2017}{\natexlab{a}})},\ \Eprint
  {http://arxiv.org/abs/1701.08157} {arXiv:1701.08157 [physics.ins-det]}
  \BibitemShut {NoStop}%
\bibitem [{\citenamefont {Kavanagh}(2018)}]{Kavanagh:2017cru}%
  \BibitemOpen
  \bibfield  {author} {\bibinfo {author} {\bibfnamefont {B.~J.}\ \bibnamefont
  {Kavanagh}},\ }\href {\doibase 10.1103/PhysRevD.97.123013} {\bibfield
  {journal} {\bibinfo  {journal} {Phys. Rev.}\ }\textbf {\bibinfo {volume}
  {D97}},\ \bibinfo {pages} {123013} (\bibinfo {year} {2018})},\ \Eprint
  {http://arxiv.org/abs/1712.04901} {arXiv:1712.04901 [hep-ph]} \BibitemShut
  {NoStop}%
\bibitem [{\citenamefont {Fornengo}\ \emph {et~al.}(2011)\citenamefont
  {Fornengo}, \citenamefont {Panci},\ and\ \citenamefont
  {Regis}}]{Fornengo:2011sz}%
  \BibitemOpen
  \bibfield  {author} {\bibinfo {author} {\bibfnamefont {N.}~\bibnamefont
  {Fornengo}}, \bibinfo {author} {\bibfnamefont {P.}~\bibnamefont {Panci}}, \
  and\ \bibinfo {author} {\bibfnamefont {M.}~\bibnamefont {Regis}},\ }\href
  {\doibase 10.1103/PhysRevD.84.115002} {\bibfield  {journal} {\bibinfo
  {journal} {Phys. Rev.}\ }\textbf {\bibinfo {volume} {D84}},\ \bibinfo {pages}
  {115002} (\bibinfo {year} {2011})},\ \Eprint {http://arxiv.org/abs/1108.4661}
  {arXiv:1108.4661 [hep-ph]} \BibitemShut {NoStop}%
\bibitem [{\citenamefont {Del~Nobile}\ \emph {et~al.}(2012)\citenamefont
  {Del~Nobile}, \citenamefont {Kouvaris}, \citenamefont {Panci}, \citenamefont
  {Sannino},\ and\ \citenamefont {Virkajarvi}}]{DelNobile:2012tx}%
  \BibitemOpen
  \bibfield  {author} {\bibinfo {author} {\bibfnamefont {E.}~\bibnamefont
  {Del~Nobile}}, \bibinfo {author} {\bibfnamefont {C.}~\bibnamefont
  {Kouvaris}}, \bibinfo {author} {\bibfnamefont {P.}~\bibnamefont {Panci}},
  \bibinfo {author} {\bibfnamefont {F.}~\bibnamefont {Sannino}}, \ and\
  \bibinfo {author} {\bibfnamefont {J.}~\bibnamefont {Virkajarvi}},\ }\href
  {\doibase 10.1088/1475-7516/2012/08/010} {\bibfield  {journal} {\bibinfo
  {journal} {JCAP}\ }\textbf {\bibinfo {volume} {1208}},\ \bibinfo {pages}
  {010} (\bibinfo {year} {2012})},\ \Eprint {http://arxiv.org/abs/1203.6652}
  {arXiv:1203.6652 [hep-ph]} \BibitemShut {NoStop}%
\bibitem [{\citenamefont {Kahlhoefer}\ \emph {et~al.}(2017)\citenamefont
  {Kahlhoefer}, \citenamefont {Kulkarni},\ and\ \citenamefont
  {Wild}}]{Kahlhoefer:2017ddj}%
  \BibitemOpen
  \bibfield  {author} {\bibinfo {author} {\bibfnamefont {F.}~\bibnamefont
  {Kahlhoefer}}, \bibinfo {author} {\bibfnamefont {S.}~\bibnamefont
  {Kulkarni}}, \ and\ \bibinfo {author} {\bibfnamefont {S.}~\bibnamefont
  {Wild}},\ }\href {\doibase 10.1088/1475-7516/2017/11/016} {\bibfield
  {journal} {\bibinfo  {journal} {JCAP}\ }\textbf {\bibinfo {volume} {1711}},\
  \bibinfo {pages} {016} (\bibinfo {year} {2017})},\ \Eprint
  {http://arxiv.org/abs/1707.08571} {arXiv:1707.08571 [hep-ph]} \BibitemShut
  {NoStop}%
\bibitem [{\citenamefont {Fan}\ \emph {et~al.}(2010)\citenamefont {Fan},
  \citenamefont {Reece},\ and\ \citenamefont {Wang}}]{Fan:2010gt}%
  \BibitemOpen
  \bibfield  {author} {\bibinfo {author} {\bibfnamefont {J.}~\bibnamefont
  {Fan}}, \bibinfo {author} {\bibfnamefont {M.}~\bibnamefont {Reece}}, \ and\
  \bibinfo {author} {\bibfnamefont {L.-T.}\ \bibnamefont {Wang}},\ }\href
  {\doibase 10.1088/1475-7516/2010/11/042} {\bibfield  {journal} {\bibinfo
  {journal} {JCAP}\ }\textbf {\bibinfo {volume} {1011}},\ \bibinfo {pages}
  {042} (\bibinfo {year} {2010})},\ \Eprint {http://arxiv.org/abs/1008.1591}
  {arXiv:1008.1591 [hep-ph]} \BibitemShut {NoStop}%
\bibitem [{\citenamefont {Fitzpatrick}\ \emph {et~al.}(2013)\citenamefont
  {Fitzpatrick}, \citenamefont {Haxton}, \citenamefont {Katz}, \citenamefont
  {Lubbers},\ and\ \citenamefont {Xu}}]{Fitzpatrick:2012ix}%
  \BibitemOpen
  \bibfield  {author} {\bibinfo {author} {\bibfnamefont {A.~L.}\ \bibnamefont
  {Fitzpatrick}}, \bibinfo {author} {\bibfnamefont {W.}~\bibnamefont {Haxton}},
  \bibinfo {author} {\bibfnamefont {E.}~\bibnamefont {Katz}}, \bibinfo {author}
  {\bibfnamefont {N.}~\bibnamefont {Lubbers}}, \ and\ \bibinfo {author}
  {\bibfnamefont {Y.}~\bibnamefont {Xu}},\ }\href {\doibase
  10.1088/1475-7516/2013/02/004} {\bibfield  {journal} {\bibinfo  {journal}
  {JCAP}\ }\textbf {\bibinfo {volume} {1302}},\ \bibinfo {pages} {004}
  (\bibinfo {year} {2013})},\ \Eprint {http://arxiv.org/abs/1203.3542}
  {arXiv:1203.3542 [hep-ph]} \BibitemShut {NoStop}%
\bibitem [{\citenamefont {Cirelli}\ \emph {et~al.}(2013)\citenamefont
  {Cirelli}, \citenamefont {Del~Nobile},\ and\ \citenamefont
  {Panci}}]{DelNobile:2013sia}%
  \BibitemOpen
  \bibfield  {author} {\bibinfo {author} {\bibfnamefont {M.}~\bibnamefont
  {Cirelli}}, \bibinfo {author} {\bibfnamefont {E.}~\bibnamefont {Del~Nobile}},
  \ and\ \bibinfo {author} {\bibfnamefont {P.}~\bibnamefont {Panci}},\ }\href
  {\doibase 10.1088/1475-7516/2013/10/019} {\bibfield  {journal} {\bibinfo
  {journal} {JCAP}\ }\textbf {\bibinfo {volume} {1310}},\ \bibinfo {pages}
  {019} (\bibinfo {year} {2013})},\ \Eprint {http://arxiv.org/abs/1307.5955}
  {arXiv:1307.5955 [hep-ph]} \BibitemShut {NoStop}%
\bibitem [{\citenamefont {Del~Nobile}(2018)}]{DelNobile:2018dfg}%
  \BibitemOpen
  \bibfield  {author} {\bibinfo {author} {\bibfnamefont {E.}~\bibnamefont
  {Del~Nobile}},\ }\href {\doibase 10.1103/PhysRevD.98.123003} {\bibfield
  {journal} {\bibinfo  {journal} {Phys. Rev.}\ }\textbf {\bibinfo {volume}
  {D98}},\ \bibinfo {pages} {123003} (\bibinfo {year} {2018})},\ \Eprint
  {http://arxiv.org/abs/1806.01291} {arXiv:1806.01291 [hep-ph]} \BibitemShut
  {NoStop}%
\bibitem [{\citenamefont {Catena}\ \emph {et~al.}(2019)\citenamefont {Catena},
  \citenamefont {Fridell},\ and\ \citenamefont {Krauss}}]{Catena:2019hzw}%
  \BibitemOpen
  \bibfield  {author} {\bibinfo {author} {\bibfnamefont {R.}~\bibnamefont
  {Catena}}, \bibinfo {author} {\bibfnamefont {K.}~\bibnamefont {Fridell}}, \
  and\ \bibinfo {author} {\bibfnamefont {M.~B.}\ \bibnamefont {Krauss}},\
  }\href {\doibase 10.1007/JHEP08(2019)030} {\bibfield  {journal} {\bibinfo
  {journal} {JHEP}\ }\textbf {\bibinfo {volume} {08}},\ \bibinfo {pages} {030}
  (\bibinfo {year} {2019})},\ \Eprint {http://arxiv.org/abs/1907.02910}
  {arXiv:1907.02910 [hep-ph]} \BibitemShut {NoStop}%
\bibitem [{\citenamefont {Essig}\ \emph {et~al.}(2016)\citenamefont {Essig},
  \citenamefont {Fernandez-Serra}, \citenamefont {Mardon}, \citenamefont
  {Soto}, \citenamefont {Volansky},\ and\ \citenamefont {Yu}}]{Essig:2015cda}%
  \BibitemOpen
  \bibfield  {author} {\bibinfo {author} {\bibfnamefont {R.}~\bibnamefont
  {Essig}}, \bibinfo {author} {\bibfnamefont {M.}~\bibnamefont
  {Fernandez-Serra}}, \bibinfo {author} {\bibfnamefont {J.}~\bibnamefont
  {Mardon}}, \bibinfo {author} {\bibfnamefont {A.}~\bibnamefont {Soto}},
  \bibinfo {author} {\bibfnamefont {T.}~\bibnamefont {Volansky}}, \ and\
  \bibinfo {author} {\bibfnamefont {T.-T.}\ \bibnamefont {Yu}},\ }\href
  {\doibase 10.1007/JHEP05(2016)046} {\bibfield  {journal} {\bibinfo  {journal}
  {JHEP}\ }\textbf {\bibinfo {volume} {05}},\ \bibinfo {pages} {046} (\bibinfo
  {year} {2016})},\ \Eprint {http://arxiv.org/abs/1509.01598} {arXiv:1509.01598
  [hep-ph]} \BibitemShut {NoStop}%
\bibitem [{\citenamefont {Derenzo}\ \emph {et~al.}(2017)\citenamefont
  {Derenzo}, \citenamefont {Essig}, \citenamefont {Massari}, \citenamefont
  {Soto},\ and\ \citenamefont {Yu}}]{Derenzo:2016fse}%
  \BibitemOpen
  \bibfield  {author} {\bibinfo {author} {\bibfnamefont {S.}~\bibnamefont
  {Derenzo}}, \bibinfo {author} {\bibfnamefont {R.}~\bibnamefont {Essig}},
  \bibinfo {author} {\bibfnamefont {A.}~\bibnamefont {Massari}}, \bibinfo
  {author} {\bibfnamefont {A.}~\bibnamefont {Soto}}, \ and\ \bibinfo {author}
  {\bibfnamefont {T.-T.}\ \bibnamefont {Yu}},\ }\href {\doibase
  10.1103/PhysRevD.96.016026} {\bibfield  {journal} {\bibinfo  {journal} {Phys.
  Rev.}\ }\textbf {\bibinfo {volume} {D96}},\ \bibinfo {pages} {016026}
  (\bibinfo {year} {2017})},\ \Eprint {http://arxiv.org/abs/1607.01009}
  {arXiv:1607.01009 [hep-ph]} \BibitemShut {NoStop}%
\bibitem [{\citenamefont {Zaharijas}\ and\ \citenamefont
  {Farrar}(2005)}]{Zaharijas:2004jv}%
  \BibitemOpen
  \bibfield  {author} {\bibinfo {author} {\bibfnamefont {G.}~\bibnamefont
  {Zaharijas}}\ and\ \bibinfo {author} {\bibfnamefont {G.~R.}\ \bibnamefont
  {Farrar}},\ }\href {\doibase 10.1103/PhysRevD.72.083502} {\bibfield
  {journal} {\bibinfo  {journal} {Phys. Rev.}\ }\textbf {\bibinfo {volume}
  {D72}},\ \bibinfo {pages} {083502} (\bibinfo {year} {2005})},\ \Eprint
  {http://arxiv.org/abs/astro-ph/0406531} {arXiv:astro-ph/0406531 [astro-ph]}
  \BibitemShut {NoStop}%
\bibitem [{\citenamefont {Emken}\ \emph {et~al.}(2017)\citenamefont {Emken},
  \citenamefont {Kouvaris},\ and\ \citenamefont {Shoemaker}}]{Emken:2017erx}%
  \BibitemOpen
  \bibfield  {author} {\bibinfo {author} {\bibfnamefont {T.}~\bibnamefont
  {Emken}}, \bibinfo {author} {\bibfnamefont {C.}~\bibnamefont {Kouvaris}}, \
  and\ \bibinfo {author} {\bibfnamefont {I.~M.}\ \bibnamefont {Shoemaker}},\
  }\href {\doibase 10.1103/PhysRevD.96.015018} {\bibfield  {journal} {\bibinfo
  {journal} {Phys. Rev.}\ }\textbf {\bibinfo {volume} {D96}},\ \bibinfo {pages}
  {015018} (\bibinfo {year} {2017})},\ \Eprint
  {http://arxiv.org/abs/1702.07750} {arXiv:1702.07750 [hep-ph]} \BibitemShut
  {NoStop}%
\bibitem [{\citenamefont {Mahdawi}\ and\ \citenamefont
  {Farrar}(2017)}]{Mahdawi:2017cxz}%
  \BibitemOpen
  \bibfield  {author} {\bibinfo {author} {\bibfnamefont {M.~S.}\ \bibnamefont
  {Mahdawi}}\ and\ \bibinfo {author} {\bibfnamefont {G.~R.}\ \bibnamefont
  {Farrar}},\ }\href {\doibase 10.1088/1475-7516/2017/12/004} {\bibfield
  {journal} {\bibinfo  {journal} {JCAP}\ }\textbf {\bibinfo {volume} {1712}},\
  \bibinfo {pages} {004} (\bibinfo {year} {2017})},\ \Eprint
  {http://arxiv.org/abs/1709.00430} {arXiv:1709.00430 [hep-ph]} \BibitemShut
  {NoStop}%
\bibitem [{\citenamefont {Emken}\ and\ \citenamefont
  {Kouvaris}(2018)}]{Emken:2018run}%
  \BibitemOpen
  \bibfield  {author} {\bibinfo {author} {\bibfnamefont {T.}~\bibnamefont
  {Emken}}\ and\ \bibinfo {author} {\bibfnamefont {C.}~\bibnamefont
  {Kouvaris}},\ }\href {\doibase 10.1103/PhysRevD.97.115047} {\bibfield
  {journal} {\bibinfo  {journal} {Phys. Rev.}\ }\textbf {\bibinfo {volume}
  {D97}},\ \bibinfo {pages} {115047} (\bibinfo {year} {2018})},\ \Eprint
  {http://arxiv.org/abs/1802.04764} {arXiv:1802.04764 [hep-ph]} \BibitemShut
  {NoStop}%
\bibitem [{\citenamefont {Mahdawi}\ and\ \citenamefont
  {Farrar}(2018)}]{Mahdawi:2018euy}%
  \BibitemOpen
  \bibfield  {author} {\bibinfo {author} {\bibfnamefont {M.~S.}\ \bibnamefont
  {Mahdawi}}\ and\ \bibinfo {author} {\bibfnamefont {G.~R.}\ \bibnamefont
  {Farrar}},\ }\href {\doibase 10.1088/1475-7516/2018/10/007} {\bibfield
  {journal} {\bibinfo  {journal} {JCAP}\ }\textbf {\bibinfo {volume} {1810}},\
  \bibinfo {pages} {007} (\bibinfo {year} {2018})},\ \Eprint
  {http://arxiv.org/abs/1804.03073} {arXiv:1804.03073 [hep-ph]} \BibitemShut
  {NoStop}%
\bibitem [{\citenamefont {Emken}\ \emph {et~al.}(2019)\citenamefont {Emken},
  \citenamefont {Essig}, \citenamefont {Kouvaris},\ and\ \citenamefont
  {Sholapurkar}}]{Emken:2019tni}%
  \BibitemOpen
  \bibfield  {author} {\bibinfo {author} {\bibfnamefont {T.}~\bibnamefont
  {Emken}}, \bibinfo {author} {\bibfnamefont {R.}~\bibnamefont {Essig}},
  \bibinfo {author} {\bibfnamefont {C.}~\bibnamefont {Kouvaris}}, \ and\
  \bibinfo {author} {\bibfnamefont {M.}~\bibnamefont {Sholapurkar}},\ }\href
  {\doibase 10.1088/1475-7516/2019/09/070} {\bibfield  {journal} {\bibinfo
  {journal} {JCAP}\ }\textbf {\bibinfo {volume} {1909}},\ \bibinfo {pages}
  {070} (\bibinfo {year} {2019})},\ \Eprint {http://arxiv.org/abs/1905.06348}
  {arXiv:1905.06348 [hep-ph]} \BibitemShut {NoStop}%
\bibitem [{\citenamefont {Albuquerque}\ and\ \citenamefont
  {Baudis}(2003)}]{Albuquerque:2003ei}%
  \BibitemOpen
  \bibfield  {author} {\bibinfo {author} {\bibfnamefont {I.~F.~M.}\
  \bibnamefont {Albuquerque}}\ and\ \bibinfo {author} {\bibfnamefont
  {L.}~\bibnamefont {Baudis}},\ }\href {\doibase 10.1103/PhysRevLett.90.221301,
  10.1103/PhysRevLett.91.229903} {\bibfield  {journal} {\bibinfo  {journal}
  {Phys. Rev. Lett.}\ }\textbf {\bibinfo {volume} {90}},\ \bibinfo {pages}
  {221301} (\bibinfo {year} {2003})},\ \bibinfo {note} {[Erratum: Phys. Rev.
  Lett.91,229903(2003)]},\ \Eprint {http://arxiv.org/abs/astro-ph/0301188}
  {arXiv:astro-ph/0301188 [astro-ph]} \BibitemShut {NoStop}%
\bibitem [{\citenamefont {Kouvaris}\ and\ \citenamefont
  {Shoemaker}(2014)}]{Kouvaris:2014lpa}%
  \BibitemOpen
  \bibfield  {author} {\bibinfo {author} {\bibfnamefont {C.}~\bibnamefont
  {Kouvaris}}\ and\ \bibinfo {author} {\bibfnamefont {I.~M.}\ \bibnamefont
  {Shoemaker}},\ }\href {\doibase 10.1103/PhysRevD.90.095011} {\bibfield
  {journal} {\bibinfo  {journal} {Phys. Rev.}\ }\textbf {\bibinfo {volume}
  {D90}},\ \bibinfo {pages} {095011} (\bibinfo {year} {2014})},\ \Eprint
  {http://arxiv.org/abs/1405.1729} {arXiv:1405.1729 [hep-ph]} \BibitemShut
  {NoStop}%
\bibitem [{\citenamefont {Kouvaris}(2016)}]{Kouvaris:2015laa}%
  \BibitemOpen
  \bibfield  {author} {\bibinfo {author} {\bibfnamefont {C.}~\bibnamefont
  {Kouvaris}},\ }\href {\doibase 10.1103/PhysRevD.93.035023} {\bibfield
  {journal} {\bibinfo  {journal} {Phys. Rev.}\ }\textbf {\bibinfo {volume}
  {D93}},\ \bibinfo {pages} {035023} (\bibinfo {year} {2016})},\ \Eprint
  {http://arxiv.org/abs/1509.08720} {arXiv:1509.08720 [hep-ph]} \BibitemShut
  {NoStop}%
\bibitem [{\citenamefont {Davis}(2017)}]{Davis:2017noy}%
  \BibitemOpen
  \bibfield  {author} {\bibinfo {author} {\bibfnamefont {J.~H.}\ \bibnamefont
  {Davis}},\ }\href {\doibase 10.1103/PhysRevLett.119.211302} {\bibfield
  {journal} {\bibinfo  {journal} {Phys. Rev. Lett.}\ }\textbf {\bibinfo
  {volume} {119}},\ \bibinfo {pages} {211302} (\bibinfo {year} {2017})},\
  \Eprint {http://arxiv.org/abs/1708.01484} {arXiv:1708.01484 [hep-ph]}
  \BibitemShut {NoStop}%
\bibitem [{\citenamefont {Hooper}\ and\ \citenamefont
  {McDermott}(2018)}]{Hooper:2018bfw}%
  \BibitemOpen
  \bibfield  {author} {\bibinfo {author} {\bibfnamefont {D.}~\bibnamefont
  {Hooper}}\ and\ \bibinfo {author} {\bibfnamefont {S.~D.}\ \bibnamefont
  {McDermott}},\ }\href {\doibase 10.1103/PhysRevD.97.115006} {\bibfield
  {journal} {\bibinfo  {journal} {Phys. Rev.}\ }\textbf {\bibinfo {volume}
  {D97}},\ \bibinfo {pages} {115006} (\bibinfo {year} {2018})},\ \Eprint
  {http://arxiv.org/abs/1802.03025} {arXiv:1802.03025 [hep-ph]} \BibitemShut
  {NoStop}%
\bibitem [{\citenamefont {Bramante}\ \emph
  {et~al.}(2019{\natexlab{a}})\citenamefont {Bramante}, \citenamefont
  {Broerman}, \citenamefont {Kumar}, \citenamefont {Lang}, \citenamefont
  {Pospelov},\ and\ \citenamefont {Raj}}]{Bramante:2018tos}%
  \BibitemOpen
  \bibfield  {author} {\bibinfo {author} {\bibfnamefont {J.}~\bibnamefont
  {Bramante}}, \bibinfo {author} {\bibfnamefont {B.}~\bibnamefont {Broerman}},
  \bibinfo {author} {\bibfnamefont {J.}~\bibnamefont {Kumar}}, \bibinfo
  {author} {\bibfnamefont {R.~F.}\ \bibnamefont {Lang}}, \bibinfo {author}
  {\bibfnamefont {M.}~\bibnamefont {Pospelov}}, \ and\ \bibinfo {author}
  {\bibfnamefont {N.}~\bibnamefont {Raj}},\ }\href {\doibase
  10.1103/PhysRevD.99.083010} {\bibfield  {journal} {\bibinfo  {journal} {Phys.
  Rev.}\ }\textbf {\bibinfo {volume} {D99}},\ \bibinfo {pages} {083010}
  (\bibinfo {year} {2019}{\natexlab{a}})},\ \Eprint
  {http://arxiv.org/abs/1812.09325} {arXiv:1812.09325 [hep-ph]} \BibitemShut
  {NoStop}%
\bibitem [{\citenamefont {Bramante}\ \emph
  {et~al.}(2019{\natexlab{b}})\citenamefont {Bramante}, \citenamefont {Kumar},\
  and\ \citenamefont {Raj}}]{Bramante:2019yss}%
  \BibitemOpen
  \bibfield  {author} {\bibinfo {author} {\bibfnamefont {J.}~\bibnamefont
  {Bramante}}, \bibinfo {author} {\bibfnamefont {J.}~\bibnamefont {Kumar}}, \
  and\ \bibinfo {author} {\bibfnamefont {N.}~\bibnamefont {Raj}},\ }\href
  {\doibase 10.1103/PhysRevD.100.123016} {\bibfield  {journal} {\bibinfo
  {journal} {Phys. Rev.}\ }\textbf {\bibinfo {volume} {D100}},\ \bibinfo
  {pages} {123016} (\bibinfo {year} {2019}{\natexlab{b}})},\ \Eprint
  {http://arxiv.org/abs/1910.05380} {arXiv:1910.05380 [hep-ph]} \BibitemShut
  {NoStop}%
\bibitem [{\citenamefont {Emken}(2019)}]{Emken:2019hgy}%
  \BibitemOpen
  \bibfield  {author} {\bibinfo {author} {\bibfnamefont {T.}~\bibnamefont
  {Emken}},\ }\emph {\bibinfo {title} {{Dark Matter in the Earth and the Sun --
  Simulating Underground Scatterings for the Direct Detection of Low-Mass Dark
  Matter}}},\ \href@noop {} {Ph.D. thesis},\ \bibinfo  {school} {University of
  Southern Denmark, $\text{CP}^\text{3}$-Origins} (\bibinfo {year} {2019}),\
  \Eprint {http://arxiv.org/abs/1906.07541} {arXiv:1906.07541 [hep-ph]}
  \BibitemShut {NoStop}%
\bibitem [{\citenamefont {Dziewonski}\ and\ \citenamefont
  {Anderson}(1981)}]{Dziewonski:1981xy}%
  \BibitemOpen
  \bibfield  {author} {\bibinfo {author} {\bibfnamefont {A.~M.}\ \bibnamefont
  {Dziewonski}}\ and\ \bibinfo {author} {\bibfnamefont {D.~L.}\ \bibnamefont
  {Anderson}},\ }\href {\doibase 10.1016/0031-9201(81)90046-7} {\bibfield
  {journal} {\bibinfo  {journal} {Phys. Earth Planet. Interiors}\ }\textbf
  {\bibinfo {volume} {25}},\ \bibinfo {pages} {297} (\bibinfo {year}
  {1981})}\BibitemShut {NoStop}%
\bibitem [{\citenamefont {{McDonough}}(2003)}]{McDonough:2003}%
  \BibitemOpen
  \bibfield  {author} {\bibinfo {author} {\bibfnamefont {W.~F.}\ \bibnamefont
  {{McDonough}}},\ }\href {\doibase 10.1016/B0-08-043751-6/02015-6} {\bibfield
  {journal} {\bibinfo  {journal} {Treatise on Geochemistry}\ }\textbf {\bibinfo
  {volume} {2}},\ \bibinfo {pages} {568} (\bibinfo {year} {2003})}\BibitemShut
  {NoStop}%
\bibitem [{\citenamefont {Cowan}\ \emph {et~al.}(2011)\citenamefont {Cowan},
  \citenamefont {Cranmer}, \citenamefont {Gross},\ and\ \citenamefont
  {Vitells}}]{Cowan:2010js}%
  \BibitemOpen
  \bibfield  {author} {\bibinfo {author} {\bibfnamefont {G.}~\bibnamefont
  {Cowan}}, \bibinfo {author} {\bibfnamefont {K.}~\bibnamefont {Cranmer}},
  \bibinfo {author} {\bibfnamefont {E.}~\bibnamefont {Gross}}, \ and\ \bibinfo
  {author} {\bibfnamefont {O.}~\bibnamefont {Vitells}},\ }\href {\doibase
  10.1140/epjc/s10052-011-1554-0, 10.1140/epjc/s10052-013-2501-z} {\bibfield
  {journal} {\bibinfo  {journal} {Eur. Phys. J.}\ }\textbf {\bibinfo {volume}
  {C71}},\ \bibinfo {pages} {1554} (\bibinfo {year} {2011})},\ \bibinfo {note}
  {[Erratum: Eur. Phys. J.C73,2501(2013)]},\ \Eprint
  {http://arxiv.org/abs/1007.1727} {arXiv:1007.1727 [physics.data-an]}
  \BibitemShut {NoStop}%
\bibitem [{\citenamefont {Armengaud}\ \emph {et~al.}(2019)\citenamefont
  {Armengaud} \emph {et~al.}}]{Armengaud:2019kfj}%
  \BibitemOpen
  \bibfield  {author} {\bibinfo {author} {\bibfnamefont {E.}~\bibnamefont
  {Armengaud}} \emph {et~al.} (\bibinfo {collaboration} {EDELWEISS}),\ }\href
  {\doibase 10.1103/PhysRevD.99.082003} {\bibfield  {journal} {\bibinfo
  {journal} {Phys. Rev.}\ }\textbf {\bibinfo {volume} {D99}},\ \bibinfo {pages}
  {082003} (\bibinfo {year} {2019})},\ \Eprint
  {http://arxiv.org/abs/1901.03588} {arXiv:1901.03588 [astro-ph.GA]}
  \BibitemShut {NoStop}%
\bibitem [{\citenamefont {Arnaud}\ \emph {et~al.}(2020)\citenamefont {Arnaud}
  \emph {et~al.}}]{Arnaud:2020svb}%
  \BibitemOpen
  \bibfield  {author} {\bibinfo {author} {\bibfnamefont {Q.}~\bibnamefont
  {Arnaud}} \emph {et~al.} (\bibinfo {collaboration} {EDELWEISS}),\ }\href@noop
  {} {\  (\bibinfo {year} {2020})},\ \Eprint {http://arxiv.org/abs/2003.01046}
  {arXiv:2003.01046 [astro-ph.GA]} \BibitemShut {NoStop}%
\bibitem [{\citenamefont {Strauss}\ \emph {et~al.}(2017)\citenamefont {Strauss}
  \emph {et~al.}}]{Strauss:2017cam}%
  \BibitemOpen
  \bibfield  {author} {\bibinfo {author} {\bibfnamefont {R.}~\bibnamefont
  {Strauss}} \emph {et~al.},\ }\href {\doibase 10.1103/PhysRevD.96.022009}
  {\bibfield  {journal} {\bibinfo  {journal} {Phys. Rev. D}\ }\textbf {\bibinfo
  {volume} {96}},\ \bibinfo {pages} {022009} (\bibinfo {year} {2017})},\
  \Eprint {http://arxiv.org/abs/1704.04317} {arXiv:1704.04317
  [physics.ins-det]} \BibitemShut {NoStop}%
\bibitem [{\citenamefont {Angloher}\ \emph
  {et~al.}(2017{\natexlab{b}})\citenamefont {Angloher} \emph
  {et~al.}}]{Angloher:2017sxg}%
  \BibitemOpen
  \bibfield  {author} {\bibinfo {author} {\bibfnamefont {G.}~\bibnamefont
  {Angloher}} \emph {et~al.} (\bibinfo {collaboration} {CRESST}),\ }\href
  {\doibase 10.1140/epjc/s10052-017-5223-9} {\bibfield  {journal} {\bibinfo
  {journal} {Eur. Phys. J. C}\ }\textbf {\bibinfo {volume} {77}},\ \bibinfo
  {pages} {637} (\bibinfo {year} {2017}{\natexlab{b}})},\ \Eprint
  {http://arxiv.org/abs/1707.06749} {arXiv:1707.06749 [astro-ph.CO]}
  \BibitemShut {NoStop}%
\bibitem [{\citenamefont {Abdelhameed}\ \emph {et~al.}(2019)\citenamefont
  {Abdelhameed} \emph {et~al.}}]{Abdelhameed:2019hmk}%
  \BibitemOpen
  \bibfield  {author} {\bibinfo {author} {\bibfnamefont {A.~H.}\ \bibnamefont
  {Abdelhameed}} \emph {et~al.} (\bibinfo {collaboration} {CRESST}),\ }\href
  {\doibase 10.1103/PhysRevD.100.102002} {\bibfield  {journal} {\bibinfo
  {journal} {Phys. Rev. D}\ }\textbf {\bibinfo {volume} {100}},\ \bibinfo
  {pages} {102002} (\bibinfo {year} {2019})},\ \Eprint
  {http://arxiv.org/abs/1904.00498} {arXiv:1904.00498 [astro-ph.CO]}
  \BibitemShut {NoStop}%
\bibitem [{\citenamefont {Cappiello}\ and\ \citenamefont
  {Beacom}(2019)}]{Cappiello:2019qsw}%
  \BibitemOpen
  \bibfield  {author} {\bibinfo {author} {\bibfnamefont {C.}~\bibnamefont
  {Cappiello}}\ and\ \bibinfo {author} {\bibfnamefont {J.~F.}\ \bibnamefont
  {Beacom}},\ }\href {\doibase 10.1103/PhysRevD.100.103011} {\bibfield
  {journal} {\bibinfo  {journal} {Phys. Rev. D}\ }\textbf {\bibinfo {volume}
  {100}},\ \bibinfo {pages} {103011} (\bibinfo {year} {2019})},\ \Eprint
  {http://arxiv.org/abs/1906.11283} {arXiv:1906.11283 [hep-ph]} \BibitemShut
  {NoStop}%
\bibitem [{\citenamefont {Bellotti}(1988)}]{Bellotti:1988bw}%
  \BibitemOpen
  \bibfield  {author} {\bibinfo {author} {\bibfnamefont {E.}~\bibnamefont
  {Bellotti}},\ }\href {\doibase 10.1016/0168-9002(88)91093-5} {\bibfield
  {journal} {\bibinfo  {journal} {Nucl.\ Instrum.\ Meth.\ A}\ }\textbf
  {\bibinfo {volume} {264}},\ \bibinfo {pages} {1} (\bibinfo {year}
  {1988})}\BibitemShut {NoStop}%
\bibitem [{\citenamefont {Urquijo}(2016)}]{Urquijo:2016dxd}%
  \BibitemOpen
  \bibfield  {author} {\bibinfo {author} {\bibfnamefont {P.}~\bibnamefont
  {Urquijo}},\ }\bibfield  {booktitle} {\emph {\bibinfo {booktitle}
  {{Proceedings, Heavy Ion Accelerator Symposium on Fundamental and Applied
  Science (HIAS 2015): Canberra, Australia, September 14-18, 2015}}},\ }\href
  {\doibase 10.1051/epjconf/201612304002} {\bibfield  {journal} {\bibinfo
  {journal} {EPJ Web Conf.}\ }\textbf {\bibinfo {volume} {123}},\ \bibinfo
  {pages} {04002} (\bibinfo {year} {2016})},\ \Eprint
  {http://arxiv.org/abs/1605.03299} {arXiv:1605.03299 [physics.ins-det]}
  \BibitemShut {NoStop}%
\bibitem [{\citenamefont {Oliphant}(06  )}]{numpy}%
  \BibitemOpen
  \bibfield  {author} {\bibinfo {author} {\bibfnamefont {T.}~\bibnamefont
  {Oliphant}},\ }\href {http://www.numpy.org/} {\enquote {\bibinfo {title}
  {{NumPy}: A guide to {NumPy}},}\ }\bibinfo {howpublished} {USA: Trelgol
  Publishing} (\bibinfo {year} {2006--}),\ \bibinfo {note} {[Online; accessed
  03/04/2020]}\BibitemShut {NoStop}%
\bibitem [{\citenamefont {Jones}\ \emph {et~al.}(01  )\citenamefont {Jones},
  \citenamefont {Oliphant}, \citenamefont {Peterson} \emph {et~al.}}]{scipy}%
  \BibitemOpen
  \bibfield  {author} {\bibinfo {author} {\bibfnamefont {E.}~\bibnamefont
  {Jones}}, \bibinfo {author} {\bibfnamefont {T.}~\bibnamefont {Oliphant}},
  \bibinfo {author} {\bibfnamefont {P.}~\bibnamefont {Peterson}},  \emph
  {et~al.},\ }\href {http://www.scipy.org/} {\enquote {\bibinfo {title}
  {{SciPy}: Open source scientific tools for {Python}},}\ } (\bibinfo {year}
  {2001--}),\ \bibinfo {note} {[Online; accessed 03/04/2020]}\BibitemShut
  {NoStop}%
\bibitem [{\citenamefont {Hunter}(2007)}]{Hunter:2007}%
  \BibitemOpen
  \bibfield  {author} {\bibinfo {author} {\bibfnamefont {J.~D.}\ \bibnamefont
  {Hunter}},\ }\href {\doibase 10.1109/MCSE.2007.55} {\bibfield  {journal}
  {\bibinfo  {journal} {Computing in Science \& Engineering}\ }\textbf
  {\bibinfo {volume} {9}},\ \bibinfo {pages} {90} (\bibinfo {year}
  {2007})}\BibitemShut {NoStop}%
\end{thebibliography}%

\end{document}